\documentclass{IEEEoj}

\usepackage{xurl} 
\usepackage{cite}
\usepackage{amsmath,amssymb,amsfonts}
\usepackage{algorithmic}
\usepackage{graphicx,color}
\usepackage{textcomp}
\usepackage{subcaption} 
\usepackage{float} 
\usepackage{subcaption} 
\usepackage{algorithm}

\usepackage{booktabs}
\usepackage{tabularx}

\usepackage{array,multirow}

\usepackage{pbox}

\usepackage{pifont}

\let\labelindent\relax
\usepackage{enumitem}

\usepackage[table]{xcolor} 

\usepackage{url}

\def\BibTeX{{\rm B\kern-.05em{\sc i\kern-.025em b}\kern-.08em
    T\kern-.1667em\lower.7ex\hbox{E}\kern-.125emX}}
\AtBeginDocument{\definecolor{ojcolor}{cmyk}{0.93,0.59,0.15,0.02}}
\def\OJlogo{\vspace{-4pt}\hskip-4pt\includegraphics[height=18pt]{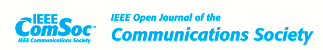}}
\begin{document}
\reviseddate{\textit{Digital Object Identifier 10.1109/OJCOMS.2026.3657212}}

\title{UAV-Mounted Aerial Relays in Military Communications: A Comprehensive Survey}

\author{\uppercase{FAisal Al-Kamali}\IEEEauthorrefmark{1,2},
\uppercase{FRANCOIS CHAN}\IEEEauthorrefmark{1,3}, \IEEEmembership{(Senior Member, IEEE)},
\uppercase{Hussein~A.~Ammar}\IEEEauthorrefmark{3}, 
\IEEEmembership{(Member, IEEE)},
\uppercase{James~H.~Bayes}\IEEEauthorrefmark{3}, 
\IEEEmembership{(Member, IEEE)},
\uppercase{CLAUDE D'AMOURS}\IEEEauthorrefmark{1}, \IEEEmembership{(Member, IEEE)}}

\affil{School of Electrical Engineering and Computer Science, University of Ottawa, Ottawa, ON, K1N 6N5, Canada}
\affil{Department of Electrical Engineering, Ibb University, Ibb, Yemen}
\affil{Department of Electrical and Computer Engineering, Royal Military College of Canada, Kingston, ON, K7K 7B4, Canada}

\corresp{CORRESPONDING AUTHOR: Faisal Al-Kamali (e-mail: faisalalkamali@gmail.com).}

\markboth{UAV-Mounted Aerial Relays in Military Communications: A Comprehensive Survey}{Al-Kamali \textit{et al.}}


\begin{abstract}
 Relays are pivotal in military communication networks, expanding coverage and ensuring reliable connectivity in challenging operational environments. While traditional terrestrial relays (TR) are constrained by fixed locations and vulnerability to physical obstructions, unmanned aerial vehicle (UAV)-mounted aerial relays (AR) offer a dynamic and flexible alternative by operating above obstacles and adapting to changing battlefield conditions. This paper provides a comprehensive survey of AR systems in military communications, presenting a detailed comparison between AR and TR paradigms and examining two specific AR technologies: active aerial relays (AAR) and aerial reconfigurable intelligent surface (ARIS) relays. The survey delves into their operation, benefits, challenges, and military applications, supported by a qualitative analysis across metrics such as coverage, flexibility, security, and cost. A novel multi-dimensional metric, the mission-critical relay effectiveness score (MCRES), is introduced as a quantitative method for evaluating relay suitability based on mission-specific weights for critical attributes like mobility, jamming resilience, deployment speed, stealth, coverage, and autonomy.  Furthermore, we present Algorithm 1, a decision-making framework that leverages the MCRES to guide the systematic selection of the optimal relay type, AR or TR, and subsequently AAR or ARIS, tailored to the unique demands of a given military scenario, such as dynamic battlefield operations, electronic warfare, or covert missions. Finally, the paper addresses current implementation challenges and outlines promising future research directions to advance the deployment of robust and resilient UAV-mounted relay systems in contested military environments.
\end{abstract}


\maketitle

\section{Introduction}
\subsection{Background}
\IEEEPARstart{I}{n} military communications, relays are essential to ensure reliable communication in challenging environments. They extend network coverage, eliminate outage gaps, and maintain connectivity where direct links are obstructed by terrain or other obstacles. Particularly in remote or rugged areas where traditional infrastructure is absent, relays provide flexible and robust communication links ~\cite{9220886}. This capability is critical for enhancing command and control, ensuring mission success. Given the stringent requirements, such as 99.999\% availability, sub-100ms latency, and secure, jam-resistant links per Department of Defense standard, relays are becoming increasingly vital as military operations grow more complex~\cite{10715496}.

Relays are classified into active and passive types. Active relays, like amplify-and-forward (AF) and decode-and-forward (DF), amplify and/or process and retransmit signals to enhance coverage and signal quality~\cite{gu2023survey}. Passive relays, such as reconfigurable intelligent surfaces (RIS), do not amplify signals; instead, they redirect them using electronically controlled passive elements~\cite{8910627}. RIS stands out for its ability to improve signal propagation and spectral efficiency while minimizing energy consumption, making it a promising technology for future communication systems~\cite{9309152}.

In wireless communication networks, relays can also be classified into terrestrial and aerial types. Terrestrial relays (TR) are fixed ground-based systems that extend coverage and provide connectivity over large areas~\cite{gu2023survey}. TRs are essential in military applications to enable communication across diverse terrains and scenarios, extend coverage area, and improve signal reliability. They are crucial for maintaining effective communication in rugged or obstructed areas where direct line-of-sight is limited, and are often used in remote or disaster-affected locations where traditional infrastructure is lacking. TRs provide localized coverage, enhance signal strength, and ensure robust connectivity between command centers and field units. Despite their benefits, they face challenges such as susceptibility to environmental factors, and resource-intensive deployment and maintenance, particularly in hostile or rapidly changing conditions \cite{gu2023survey}. 

In contrast, aerial relays (AR), mounted on unmanned aerial vehicles (UAVs), offer dynamic and flexible communication solutions by quickly adapting to changing environments and covering areas difficult for TRs to reach~\cite{10379625}, thanks to the advantages of the UAVs. TR and AR can function as active or passive relays.

\begin{figure}[t]
    \centering
    \includegraphics[width=\columnwidth]{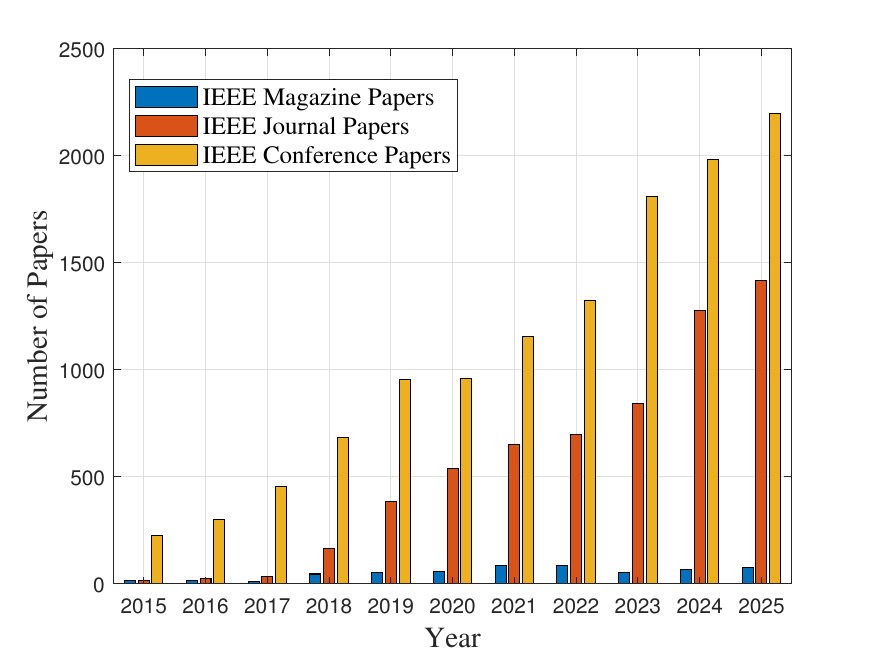}
    \caption{\textcolor{black}{Number of research papers on using UAVs in communication systems. Source: IEEE Xplore}}
    \label{fig:publications}
    \vspace{-1em}
\end{figure}

Recently, UAV-assisted wireless communication systems have emerged as a promising technology for expanding signal coverage and enhancing transmission efficiency, due to their flexible configuration~\cite{8758183, 9043702}, which allows them to adapt to rapidly changing environments. Fig.~\ref{fig:publications} illustrates the growing number of publications on the use of UAVs in communication systems, showing an increasing trend and interest in this area. In addition, the global number of UAVs, has been rising steadily year over year, fueled by rapid technological advancements and the growing adoption across military, commercial, and civilian sectors. 

In 2024, approximately 5.42 million UAVs were estimated to be in operation worldwide \cite{marketsandmarkets2024}. This number is projected to grow to around 7.51 million by 2029, reflecting a compound annual growth rate of about 6.7\% \cite{marketsandmarkets2024}. The increase is primarily driven by expanding use cases such as aerial surveillance, agriculture, infrastructure inspection, delivery services, and military operations, all of which continue to push global demand for UAVs. As a result, UAV-assisted wireless communication systems have become a major focus for researchers and industry experts, as they offer effective solutions for improving coverage, reliability, and network flexibility~\cite{9127423, 9722778, 8403626}.  
For instance, \cite{9127423} develops a machine learning framework to enhance edge computing performance in mobile vehicular environments, focusing on throughput maximization. Separately, \cite{9722778} introduces an energy efficiency optimization scheme specifically designed for multi-UAV systems supporting vehicular communications, incorporating air-to-ground (A2G) channel uncertainty modeling.

In military applications, UAVs are crucial due to their adaptability and operational advantages. Over the past decade, the military departments have rapidly expanded their unmanned systems, which now play a vital role in modern operations. The U.S. Department of Defense 2013–2038 roadmap \cite{USDoD2014} identifies UAVs as key enablers for secure, resilient communications, including use as airborne relays to extend network coverage in denied environments. They are also employed for surveillance, reconnaissance, and tactical support, delivering real-time intelligence and situational awareness \cite{USDoD2014}. Additional roles include target acquisition, precision strikes, and supply delivery in hazardous zones, minimizing risks to personnel. Their ability to operate in hostile environments and integrate with advanced military systems further improves coordination and response \cite{USDoD2014, 10715496, 9864322}.

Several recent military reports have examined various aspects of unmanned aerial systems (UAS) and their growing role in modern military operations, security challenges, and emerging technologies. According to the report \cite{ref1}, the U.S. military has taken the lead in integrating wide-area sensors, real-time battle networks, and precision-guided munitions into a reconnaissance–strike complex. UAS play a central role in this framework, supporting a wide range of missions. Report \cite{ref2} examines the characteristics, proliferation, and implications of armed UAVs for U.S. security. It concludes that, although these systems offer significant advantages in  military operations, they are not transformative weapons, and their proliferation does not warrant new arms control measures. The threat of UAV use in terrorism is discussed in report \cite{ref3}, which details their low cost, high impact potential, detection difficulties, and countermeasures. Technical and operational insights are provided in report \cite{ref4}, reviewing platforms such as the MQ-9 Reaper and RQ-4 Global Hawk, alongside cost, interoperability, and export regulation challenges. The rising demand for UAS and loitering munitions is analyzed in report \cite{ref5}, linking their prominence to recent conflicts and shifts in defense industrial bases and export policies. Finally, \cite{ref6} addresses the influence of AI-driven drone technologies on modern warfare, underscoring the ethical, legal, and security challenges they pose and advocating for international collaboration to regulate autonomous weapon systems.

These growing capabilities also highlight the importance of addressing communication limitations in UAV operations. While UAVs often maintain line-of-sight links,  obstacles like trees and buildings can block signals and degrade communication performance.  Integrating UAVs with RIS can overcome these limitations by enhancing reliability while reducing both energy consumption and system complexity compared to active relays ~\cite{9690481, 9013626, 9750059, 9684973, 8959174, 9500484, 9434412, 10261452}. There are two configurations that integrate UAVs and RIS \cite{9690481}. In the first configuration, a UAV operates with a terrestrial RIS (TRIS) mounted on building walls to reflect signals transmitted from the UAV. In the second configuration, the RIS is mounted on a UAV, creating an aerial RIS (ARIS) that reflects signals transmitted from a terrestrial base station \cite{9690481}. Considering the power limitations of small UAVs, ARIS is likely to be implemented on low-cost UAVs, where the RIS functions as a reflector and does not require a transmitter or receiver. In contrast to traditional TRIS systems, ARIS can dynamically adjust its position to optimize system performance, which has attracted the attention of more researchers in recent years~\cite{9750059, 9684973, 8959174, 9500484, 9434412}. The results in ~\cite{ 8959174} show that a communication system with ARIS outperforms one with TRIS, particularly in challenging propagation environments, by effectively optimizing UAV trajectory and passive beamforming. In~\cite{10261452}, the authors compared terrestrial base stations, aerial base stations, and aerial RIS-aided post-disaster wireless networks. Their results indicate that aerial RIS -aided networks surpass the other methods, particularly in terms of coverage.

The number of reflecting elements that can be mounted on a single UAV is constrained by the UAV’s payload capacity, battery life, and flight flexibility. Despite RIS technology being generally lightweight and having a conformal design, a single UAV-enabled ARIS might not provide the large aperture gain necessary for optimal performance. To address this issue, the authors in~\cite{9579401} proposed using multiple cooperative UAVs, or a UAV swarm, to support ARIS, as it can greatly improve system performance. However, these systems remain conceptual and are far from a complete operational solution.

Research in industrial automation provides valuable insights that can be used to improve UAV networks. The study in \cite{10558826} presents a time-sensitive network architecture using deep reinforcement learning to guarantee reliable, low-latency AI services. \cite{10409245} develops a quantum-based algorithm for efficient resource management across space, air, and ground networks. The intelligent resource adaptation work in \cite{10757329} introduces a deep learning framework for dynamic task allocation in industrial IoT systems with diverse service needs. 
\\
The acronyms used throughout the text are listed and explained Table~\ref{table:acronyms}.

\begin{table}[tbhp]
\caption{Acronyms and Abbreviations}
\label{table:acronyms}
\begin{tabular}{p{2cm}p{6cm}}
\hline
\textbf{Acronym} & \textbf{Description} \\
\hline
2AG & Air-to-ground \\
5G & Fifth generation \\
6G & Sixth generation \\
A2AD & Anti-Access/Area Denial \\
A2G & Air-to-ground \\
AAR & Active aerial relay \\
AF & Amplify-and-forward \\
AI & Artificial intelligence \\
AJP-6 & Allied joint publication 6 \\
AR & Aerial relay \\
ARES & Aerial reconfigurable embedded system \\
ARIS & Aerial reconfigurable intelligent surface \\
ATR & Aerial terrestrial relay \\
B5G & Beyond fifth generation \\
BCD & Block coordinate descent \\
C2 & Command-and-control \\
CIS & Communications and information systems \\
DEW & Directed-energy weapon \\
DF & Decode-and-forward \\
DoD & Department of Defense \\
EMS & Electromagnetic spectrum \\
EW & Electronic warfare \\
FAA & Federal Aviation Administration \\
FD & Full-duplex \\
GPS & Global Positioning System \\
HALE & High-altitude long-endurance \\
HD & Half-duplex \\
HPM & High-power microwave \\
ISR & Intelligence, surveillance, and reconnaissance \\
JCIDS & Joint Capabilities Integration and Development System \\
LPD & Low probability of detection \\
LPI & Low probability of intercept \\
MALE & Medium-altitude long-endurance \\
MCRES & Mission-critical relay effectiveness score \\
ML & Machine learning \\
MIMO & Multi-input multi-output \\
mmWave & Millimeter wave \\
NASA & National Aeronautics and Space Administration \\
NATO & North Atlantic Treaty Organization \\
RCS & Radar cross-section \\
RF & Radio frequency \\
RIS & Reconfigurable intelligent surface \\
SAR & Search and rescue \\
SARIS & Swarm-enabled ARIS \\
SEAD & Suppression of enemy air defenses \\
SNR & Signal-to-noise ratio \\
STANAG & Standardization Agreement \\
TDL & Tactical data link \\
TR & Terrestrial relay \\
TRIS & Terrestrial reconfigurable intelligent surface \\
UAS & Unmanned aerial systems \\
UAV & Unmanned aerial vehicle \\
\hline
\end{tabular}
\end{table}

\subsection{Existing Surveys}
To the best of our knowledge, no prior survey has systematically examined UAV-mounted relays in military communications. Existing surveys are predominantly civilian-focused, spanning areas such as general UAV communications, channel modeling, physical layer security, 5G/6G integration, RIS, and machine learning. While these studies do not address mission-critical or contested environments, they provide valuable foundations that can be adapted to military contexts. Accordingly, we review these civilian-oriented surveys both to extract transferable insights and to highlight the critical gaps that motivate this military-focused study. The existing literature can therefore be categorized into six main groups:
%
%
%
%
%
\subsubsection{\textbf{General UAV Communications and Deployment:}} 
Surveys such as \cite{gu2023survey, 10379625, 7470933, 7317490, 8438489, 8470897, 8660516, 8839972} examine UAV network architectures, deployment strategies, and multi-UAV coordination. These works emphasize coverage extension, reliable connectivity, and energy efficiency.
In~\cite{gu2023survey}, the authors present a detailed overview of UAV applications in communication systems. They explore the use of UAVs as aerial base stations for data collection, and power transmission, as well as relay nodes to enhance coverage and reduce interference. The paper also identifies challenges and future research directions for UAV-assisted communications. In~\cite{10379625}, the authors conduct a comprehensive survey on UAV deployment scenarios, applications, challenges, and emerging technologies in 5G and beyond networks. The study begins with a brief background, followed by a detailed classification of UAVs and a review of key studies, emphasizing both single and multiple UAV configurations.

In~\cite{7470933}, the authors provide a comprehensive overview of UAV-aided wireless communications, discussing how UAVs can be used to extend coverage, serve as relays, and distribute information. They explain the basic network architecture, channel characteristics, and key design factors for UAV communications, showing that UAVs can act as aerial relays to significantly improve the throughput, reliability, and coverage of terrestrial communication systems. Similarly,~\cite{7317490} offers a detailed survey  on UAV networks, covering topics such as UAV ad hoc mesh network characteristics, constrained routing, seamless handovers, and energy-efficient UAV operations. 

In~\cite{8438489}, the authors review design mechanisms and protocols for airborne communication networks, focusing on low-altitude, and high-altitude.
 The study in ~\cite{8470897} summarizes the benefits and challenges of cellular-connected UAVs. The authors of~\cite{8660516} discuss strategies for optimizing UAV-based wireless systems, focusing on challenges such as deployment and channel modeling. The authors of ~\cite{8839972} focus on cooperation mechanisms for cellular-connected UAV networks, recommending trajectory design and resource management protocols to improve the quality of service. 
 
\underline{\textbf{Synthesis and Military Gap:}} While these surveys offer transferable insights into UAV deployment and coordination, they remain primarily civilian-focused and do not address mission-critical requirements such as resilience under electronic warfare (EW), secure operation in contested airspaces, or integration with tactical military networks. This gap underscores the need for a dedicated study of UAV-mounted relays in military contexts.

\subsubsection{\textbf{Channel Modeling for UAV Communication:}} 
Studies in \cite{8411465, 8709739} focus on UAV channel measurements, modeling techniques, and fading characteristics, including air-to-ground (A2G) link. 
The authors of ~\cite{8411465} provide a survey on UAV communication channel modeling, discussing measurement methods, channel characteristics, and real-world challenges like airframe shadowing and non-stationarity. The study in~\cite{8709739} explores A2G propagation channel modeling for UAVs, addressing large-scale fading, small-scale fading, and multi-input multi-output (MIMO) characteristics. 

\underline{\textbf{Synthesis and Military Gap:}} Collectively, these studies establish a strong foundation for understanding UAV-specific propagation environments and provide transferable insights into channel dynamics. However, they are primarily built on benign and interference-free settings, offering little guidance for contested military environments where adversarial jamming, spoofing, and dynamic electromagnetic threats fundamentally alter channel behavior. This limitation underscores the need to extend channel modeling research toward threat-resilient and mission-critical scenarios.

\subsubsection{\textbf{Physical Layer Security in UAV Communication:}}
Works such as \cite{8883128, 9900257, 9925214} explore secrecy performance and secure communication strategies using mmWave, beamforming, massive MIMO, and NOMA. In~\cite{8883128}, the authors investigate UAV communication challenges and opportunities related to physical layer security, suggesting millimeter Wave (mmWave) communication and beamforming to boost security. The authors of~\cite{9900257} present a detailed survey on UAV physical layer security, covering both static and mobile UAV deployment scenarios, along with A2G channels. They review secrecy performance and enhancement techniques for these systems, summarize key methodologies, and outline future research challenges. In~\cite{9925214}, the authors provide a survey on UAV-aided network security, reviewing security threats, secrecy metrics, and proactive techniques using technologies like mmWave, non-orthogonal multiple access (NOMA), massive MIMO, and cognitive radio. Emerging topics such as machine learning and blockchain are also discussed for secure UAV communication.

\underline{\textbf{Synthesis and Military Gap:}} Together, these works establish a solid foundation for secure UAV communications at the physical layer, offering transferable techniques for confidentiality and resilience in benign environments. However, they largely assume civilian use cases and overlook the harsher realities of military operations, where communication must remain covert and resilient under jamming, spoofing, and EW. This gap highlights the need for research that tailors physical layer security to contested and mission-critical contexts.

\vspace{-0.5em}

\subsubsection{\textbf{UAV for 5G and 6G:}}
Several surveys \cite{10379625, 8918497, 8579209, khan2021role, 9358097, 9768113, 9598918 } analyze UAV integration into 5G and beyond 5G (B5G) networks, discussing interference management, massive MIMO, and air-to-air links. In~\cite{10379625}, the authors conduct a comprehensive survey on UAV deployment scenarios, applications, challenges, and emerging technologies in 5G and beyond networks. The study begins with a brief background, followed by a detailed classification of UAVs and a review of key studies, emphasizing both single and multiple UAV configurations. In~\cite{8918497}, a tutorial on UAV communications for B5G systems covers fundamental aspects like channel and antenna models, energy consumption, and performance metrics. \cite{8579209} presents a comprehensive survey on UAV communication in 5G and B5G. Meanwhile,~\cite{khan2021role} explores the challenges of deploying UAV relays, particularly mmWave-enabled ones, and reviews machine learning-based UAV networks and mmWave path loss models. In~\cite{9358097}, the authors survey aerial radio access networks for 6G, addressing network design, performance metrics, and technologies for energy replenishment and data transmission. The research work in~\cite{9768113} reviews UAV cellular communications for 5G, detailing solutions for interference, and coverage using massive MIMO and mmWave technologies, and discusses air-to-air communications. It also explores future B5G directions and 6G paradigms such as non-terrestrial networks, cell-free architectures, artificial intelligence (AI), RIS, and THz communication. A comprehensive survey on mmWave beamforming in UAV communications is presented in~\cite{9598918}, covering the technical potential and challenges, reviewing mmWave antenna structures and channel models, and exploring solutions for both UAV-connected mmWave cellular and mmWave-UAV ad hoc networks.  

\underline{\textbf{Synthesis and Military Gap:}} These studies underline the growing role of UAVs as integral enablers of 5G/6G networks, emphasizing advanced techniques like mmWave, massive MIMO, and RIS to enhance coverage and performance. However, their focus remains on civilian use cases, assuming cooperative environments and infrastructure support. Critical military requirements, such as resilience under jamming, survivability in contested airspaces, and secure integration with tactical data links, are not addressed, leaving a significant gap for defense-oriented research.

\subsubsection{\textbf{RIS for UAV Communication:}}
Surveys in \cite{9703337, 10208239} explore RIS-assisted UAV systems, highlighting improvements in energy efficiency, channel modeling, and IoT integration.  In~\cite{9703337}, a survey on RIS-assisted UAV systems covers optimization, communication techniques, learning methods, and highlights challenges and future directions in phase shifting, energy efficiency, federated learning, and channel modeling.  In~\cite{10208239}, the authors highlight how RIS can enhance UAV system performance, examining its integration into internet of things (IoT) models to improve energy efficiency and signal strength. The paper also addresses unresolved issues in RIS-UAV systems, and proposes new solutions. 

\underline{\textbf{Synthesis and Military Gap:}} These studies demonstrate RIS as a promising tool to enhance UAV communication performance, particularly in energy-constrained or coverage-limited scenarios. However, all existing works focus on civilian deployments under predictable conditions. The applicability of RIS-UAV systems to military missions, which require robustness against adversarial actions, environmental uncertainty, and tactical constraints, remains largely unexplored. This highlights a critical gap for military-oriented research on RIS-enabled UAV-mounted relays.

\begin{table*}[tbhp]
\caption{Existing UAV Survey Papers}
\label{table:summary_categorized}
\centering
\fontsize{9}{10}
\begin{tabular}{@{}p{0.05\textwidth} p{0.025\textwidth} p{0.12\textwidth} p{0.04\textwidth} p{0.06\textwidth} p{0.08\textwidth} p{0.48\textwidth}@{}}
\toprule
\textbf{No.} & \textbf{Year} & \textbf{Application Type} & \textbf{AR} & \textbf{ARIS} & \textbf{UAV} & \textbf{Brief Description} \\
\midrule
\multicolumn{7}{l}{\textbf{1) Civilian UAV Communications and Deployment}} \\
\cite{gu2023survey} & 2023 & Civilian & Yes & No & Yes & Reviews UAV-assisted wireless communication systems, including relays, aerial base stations, and mobile servers. \\
\cite{7470933} & 2016 & Civilian & Yes & No & Yes & Discusses UAV-aided wireless communications, architectures, and design factors. \\
\cite{7317490} & 2016 & Civilian & No & No & Yes & Reviews multi-UAV networks with focus on routing, handovers, and energy efficiency. \\
\cite{8438489} & 2018 & Civilian & No & No & Yes & Surveys airborne communication networks and challenges at different altitudes. \\
\cite{8470897} & 2019 & Civilian & Yes & No & Yes & Reviews benefits and challenges of cellular-connected UAVs. \\
\cite{8660516} & 2019 & Civilian & Yes & No & Yes & Reviews UAV-based wireless systems, deployment issues, and optimization tools. \\
\cite{8839972} & 2019 & Civilian & Yes & No & Yes & Presents cooperation protocols for UAVs in cellular networks. \\
\midrule
\multicolumn{7}{l}{\textbf{2) Channel Modeling for UAV Communication}} \\
\cite{8411465} & 2018 & Channel Modeling & No & No & Yes & Surveys UAV communication channel models, measurement methods, and airframe shadowing. \\
\cite{8709739} & 2019 & Civilian & No & No & Yes & Explores A2G propagation models including large- and small-scale fading. \\
\midrule
\multicolumn{7}{l}{\textbf{3) Physical Layer Security in UAV Communication}} \\
\cite{8883128} & 2019 & Civilian & Yes & No & Yes & Explores UAV physical layer security using mmWave and beamforming. \\
\cite{9900257} & 2022 & Civilian & Yes & Suggested & Yes & Surveys physical layer security for UAVs in static and mobile deployment. \\
\cite{9925214} & 2022 & Civilian & Yes & No & Yes & Reviews UAV network security threats, secrecy metrics, and enhancement methods. \\
\midrule
\multicolumn{7}{l}{\textbf{4) UAV for 5G and 6G}} \\
\cite{10379625} & 2024 & Civilian & Yes & No & Yes & Comprehensive survey of UAV deployment, applications, emerging technologies, and challenges in 5G/B5G. \\
\cite{8918497} & 2019 & Civilian & Yes & No & Yes & Tutorial on UAV channels, antenna models, energy, and metrics for B5G. \\
\cite{8579209} & 2019 & Civilian & Yes & No & Yes & Surveys UAV communication in 5G and B5G networks. \\
\cite{khan2021role} & 2021 & Civilian & Yes & No & Yes & Reviews UAV relays with mmWave, AI-based solutions, and security issues. \\
\cite{9358097} & 2021 & Civilian & Yes & No & Yes & Surveys aerial RAN for 6G, addressing network design, energy, and coverage. \\
\cite{9768113} & 2022 & Civilian & Yes & Yes & Yes & Reviews UAV cellular systems with massive MIMO, RIS, and air-to-air links. \\
\cite{9598918} & 2022 & Civilian & Yes & No & Yes & Surveys mmWave beamforming for UAV communication and challenges. \\
\midrule
\multicolumn{7}{l}{\textbf{5) RIS for UAV Communication}} \\
\cite{9703337} & 2022 & Civilian & No & Yes & Yes & Surveys RIS-assisted UAVs, covering optimization, energy efficiency, and learning. \\
\cite{10208239} & 2023 & Civilian & No & Yes & Yes & Reviews RIS for UAV-IoT systems, focusing on energy efficiency and open challenges. \\
\midrule
\multicolumn{7}{l}{\textbf{6) Machine Learning for UAV Communication}} \\
\cite{10283826} & 2023 & Civilian & No & No & Yes & Reviews reinforcement learning for UAV networks, autonomy, and computing. \\
\cite{10494323} & 2024 & Civilian & Yes & No & Yes & Explores ML-enhanced NOMA for UAV communications. \\
\cite{10531095} & 2024 & Civilian & Yes & No & Yes & Reviews ML techniques (federated, transfer, meta-learning) for UAV networks. \\
\midrule
\multicolumn{7}{l}{\textbf{Military-Focused Survey (This Work)}} \\
This Survey & --- & Military & Yes & Yes & Yes & Focuses on UAV and relay technologies for military communications, comparing with terrestrial relays and analyzing benefits, challenges, and future directions. \\
\bottomrule
\end{tabular}
\end{table*}

\subsubsection{\textbf{Machine Learning for UAV Communication:}}
Recent surveys \cite{10283826, 10494323, 10531095} investigate reinforcement learning, federated learning, transfer learning, and AI-driven trajectory optimization. These methods enhance adaptability and automation of the UAV. The authors of~\cite{10283826} review the use of reinforcement learning in autonomous mobile wireless networks, covering applications such as data access, resource management, UAV-assisted computing, and network security. The review also highlights key challenges and areas for future research in this area. In ~\cite{10494323}, the authors explore the use of NOMA techniques with UAVs, enhanced by machine learning (ML) in wireless communication networks. The study outlines the basics of UAVs and NOMA, discusses NOMA’s role in UAV networks, and examines ML applications, concluding with future research challenges. Similarly, the study in ~\cite{10531095} conducts a comprehensive review of advanced ML strategies, covering techniques such as federated learning, transfer learning, meta-learning, and explainable AI. The study explores the application of state-of-the-art ML algorithms and discusses their potential integration into cloud and edge computing-based network architectures.

\underline{\textbf{Synthesis and Military Gap:}} ML  and AI provide critical capabilities for autonomy, decision-making, and adaptive control in UAV networks. While current research largely focuses on civilian optimization problems, their potential for military applications, such as adversarial detection, threat-aware navigation, secure decision-making, and resilience under contested conditions, remains largely unexplored. Incorporating ML and AI into military UAV operations can significantly enhance situational awareness, survivability, and mission effectiveness, addressing challenges that conventional approaches cannot.

Table~\ref{table:summary_categorized} presents a brief summary of various related surveys, with short descriptions of each. As shown in the table, these surveys generally explore UAV-aided wireless communications, covering topics such as aerial relays, RIS, machine learning, network architectures, channel modeling, and security techniques within 5G and B5G networks. However, military applications are either not addressed or only briefly mentioned.

\vspace{-0.5em}

\subsection{Critical Insights and Research Gaps in UAV Survey Literature}
 The reviewed literature collectively advances understanding of UAV-enabled wireless systems, offering valuable insights into deployment strategies, channel characteristics, security mechanisms, and integration with emerging technologies such as 5G, 6G, RIS, and machine learning. A common feature across these works is their strong focus on civilian-oriented applications, including coverage enhancement, disaster recovery, IoT support, and network optimization under cooperative and benign conditions. Key themes emphasized in the literature include energy efficiency, multi-UAV coordination, routing, handover management, and reliability. \\
Despite these contributions, there remains a notable lack of research addressing UAV operations in adversarial or contested environments. Critical aspects such as resilience against jamming and interference, stealth-preserving communication, survivability under EW, and secure operation in dynamic threat scenarios are largely unexplored. Existing studies rarely consider mission-critical constraints or operational requirements unique to military deployments, highlighting a gap that cannot be addressed solely by extrapolating civilian-focused findings.\\
Given the increasing operational importance of UAVs and aerial relays in military communications \cite{USDoD2014}, along with the growing demand for UAV-enabled networks \cite{marketsandmarkets2024}, there is a pressing need for dedicated research that systematically investigates these challenges.
This survey fills that need by not only providing a comprehensive overview of UAV-based relay systems but also by delivering a qualitative comparison with terrestrial relays and introducing a novel quantitative framework. Our analysis yields vital insights into the design, performance, security, and optimal deployment of these systems to ensure robust connectivity in hostile and mission-critical environments.

\subsection{Aims, Contributions and Organization}
  
\subsubsection{Aims}

This paper aims to provide a comprehensive survey and an in-depth analysis of UAV and relays in military communications.

\begin{figure*}[t]
    \centering
    \includegraphics[width=\textwidth]{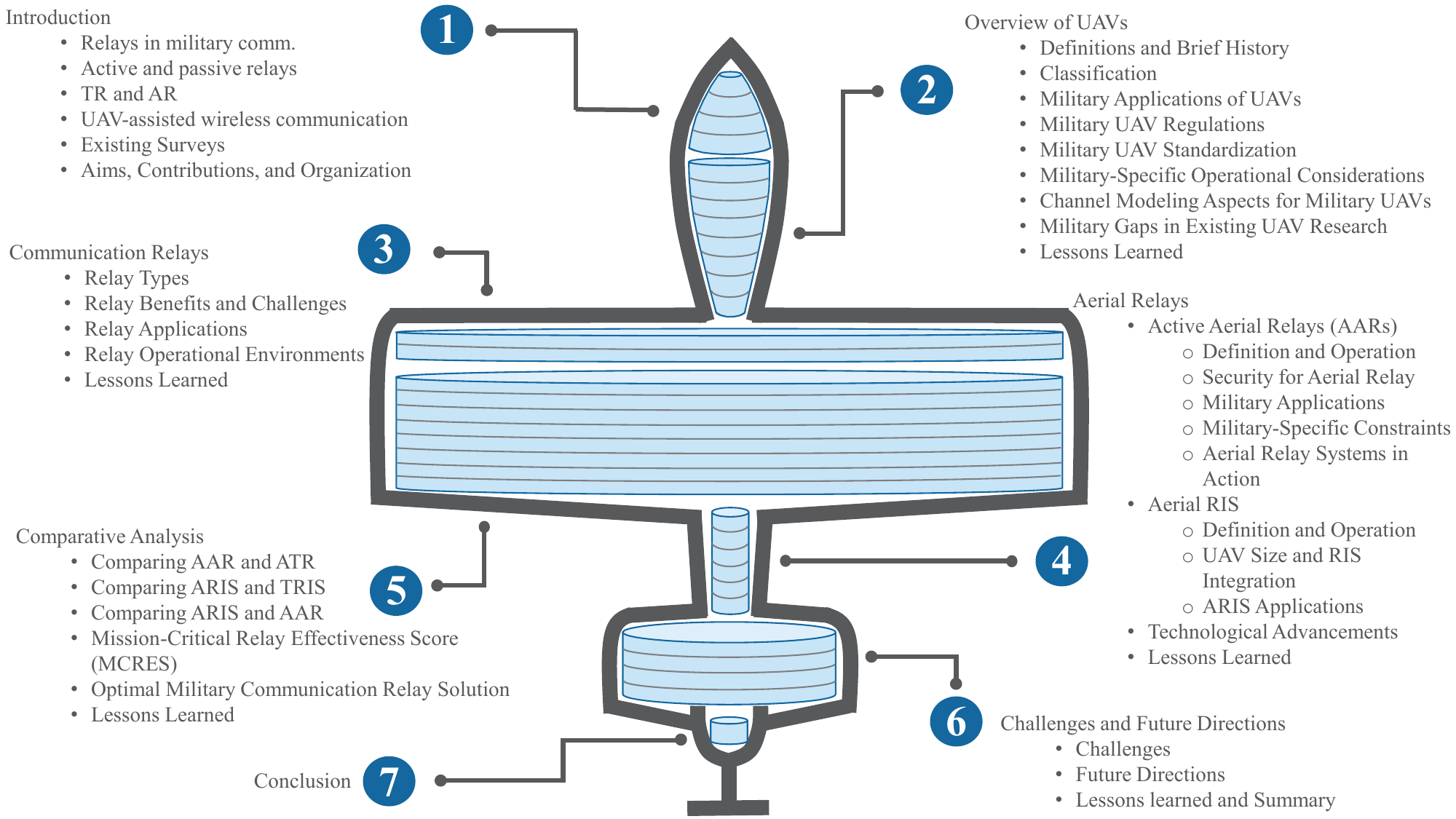}
    \caption{\textcolor{black}{Structure of the paper: The labeled  height of each green cylinder approximates the length of its corresponding section.}} 
    \label{fig:Paper_structure}
    \vspace{-1em}
\end{figure*}

\subsubsection{Contributions}
To the best of our knowledge, this is the first comprehensive survey dedicated to UAV-mounted aerial relays in military communications. The main contributions of this paper are:
\begin{enumerate}
    \item \textit{A Military-Focused Taxonomy and Qualitative Analysis:} We present a detailed comparison between AR and TR, specifically examining two key AR technologies: active aerial relays (AARs) and ARIS relays. This analysis evaluates critical performance metrics—such as operational flexibility, security, cost, and resilience—through a military operational lens, providing a foundational understanding for system selection.

    \item \textit{Introduction of a Novel Decision-Making Metric (MCRES):} A central contribution of this work is the proposal of the mission-critical relay effectiveness score (MCRES), a novel multi-dimensional metric. The MCRES offers a quantitative framework to evaluate relay suitability by integrating mission-specific weights across six doctrine-derived parameters: mobility, jamming resilience, deployment speed, stealth \& security, coverage \& performance, and autonomy \& sustainability.

    \item \textit{A Structured Relay Selection Framework (Algorithm~1):} Building upon the MCRES, we develop a structured decision-making algorithm. This framework provides military planners with a systematic, objective, and scenario-tailored method for selecting the optimal relay type—first between AR and TR, and subsequently between active and passive based on the unique demands of a given military scenario.

    \item \textit{Synthesis of Challenges and Militarily-Relevant Research Directions:} We consolidate the current implementation challenges—including limited endurance, EW susceptibility, and coordination complexity—and outline a roadmap for future research. These directions are explicitly tailored to advance the deployment of robust, resilient, and survivable UAV-mounted relay systems in contested military environments, addressing critical gaps not covered in civilian literature.
\end{enumerate}

\subsubsection{Paper Organization}

The rest of the paper is organized as follows. Section II introduces an overview of UAV and its military applications. Section III explores communication relays. Section IV explains ARs. Section V provides a comparative analysis between all types of relays. Section VI outlines the challenges and future research directions of ARs in military communications. Finally, Section VII summarizes and concludes the paper. In Fig.~\ref{fig:Paper_structure}, we show the structure of this survey paper.

\section{Overview of UAVs}
\subsection{Definition and Brief History}

    UAVs are aircrafts that operate without an onboard pilot, either controlled remotely or autonomously via onboard computers. They widely vary in size, range, and capabilities, from small handheld units to large, long-endurance aircraft\footnote{Endurance of a UAV is the maximum flight time. It depends on many factors such as battery capacity, fuel load, weight, and aerodynamic.}.   UAVs, also known as UAS, drones or remotely piloted aircraft systems, were originally developed for military purposes. Nowadays, their range of applications has expanded, and they are increasingly used beyond traditional military roles. 
          
    The development of UAVs for military use began in 1883 with aerial photography using a kite. By 1898, UAV technology was employed in the Spanish-American War for reconnaissance. In the 1930s, the British Royal Navy developed the Queen Bee, a reusable UAV for target practice. During World War II, Germany created the V-1, an unmanned flying bomb used against civilian targets. In the 1960s and 1970s, the U.S. conducted extensive surveillance with UAVs like the Ryan Firebee.
     
     The late 20th century saw the emergence of the Scout and Pioneer UAVs, which featured live video transmission and a lightweight, glider-like design, setting the stage for modern UAVs. UAV technology advanced significantly in the 1980s and 1990s, expanding their military use in surveillance and reconnaissance. In the 2000s, UAVs' role greatly expanded, particularly in military operations. Today, UAVs are widely used in civilian sectors like aerial photography, disaster response, and agriculture, as well as in scientific research for environmental monitoring and data collection in remote areas \cite{9213967, saska2016swarm, 6290694}. Their small, affordable designs have made them accessible to a broader range of users. As UAV technology has advanced, recent research into UAV communications and networking has become increasingly comprehensive. 

A single UAV may suffice for small-scale or low-risk missions, but in military operations it presents severe limitations. Its restricted sensing, processing, and payload capacity hinder effectiveness in wide-area surveillance, EW support, or sustained command-and-control (C2) missions. More critically, a lone UAV represents a single point of failure: if compromised by jamming, spoofing, or kinetic attack, the entire mission can collapse. In contrast, multi-UAV deployments provide clear military advantages. Distributed fleets shorten operation times by dividing tasks, maintain resilience by compensating for individual losses, and enhance adaptability in contested environments \cite{8022685, S20500090, 8682048, 9973697}. Swarm-based relays, for example, can reconfigure dynamically to preserve connectivity under electronic attack or when operating in GPS-denied environments, ensuring mission continuity and robustness in hostile battlespaces.

Prior studies in civilian applications have investigated UAV deployments, swarming challenges, channel and antenna modeling, interference management, GPS redundancy, security, 5G integration, and network planning \cite{8668495, 8950036, 7497529, 8606910, 8692749, 8658076, 8756296, Banafaa2021, 8682048, 8654727, 9075221, 8325268, 8869706, 8470897, 8758988, 8998329, 8253543, 8533634, 8387981}. UAVs have also been studied as aerial base stations \cite{Shayea2022}, though challenges such as limited bandwidth, unstable connectivity, and interference remain \cite{10379625, 8660516}. While these studies provide valuable insights, they are predominantly civilian-focused and rely on assumptions—such as stable environments, reliable GPS, and limited adversarial interference—that rarely hold in military scenarios. Applying these approaches to military applications requires accounting for EW threats, GPS denial, and cyber resilience, as well as ensuring compatibility with tactical data links and survivability in contested electromagnetic environments. Thus, applying these studies in military contexts demands specialized models that address battlefield-specific requirements.

\begin{figure}[t]
    \centering
    \includegraphics[width= \columnwidth]{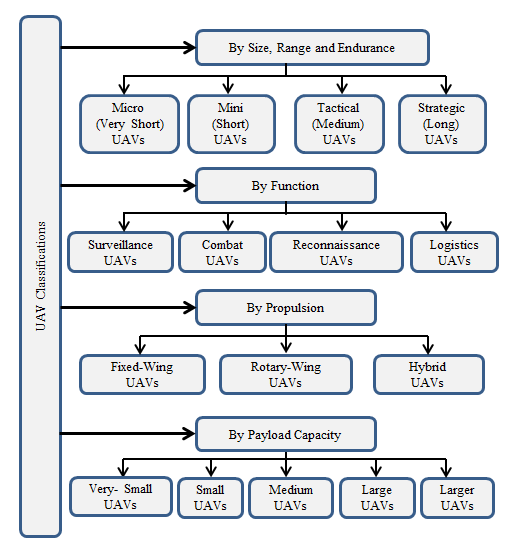} 
    \caption{\textcolor{black}{Military UAV Classifications.}} 
    \label{fig:UAV Classifications}
    \vspace{-1em}
\end{figure}

\subsection{Classification}

In this subsection, military UAVs are categorized based on size, range, function, propulsion system, and payload capacity, as illustrated in Fig.~\ref{fig:UAV Classifications} and described in detail below.

\subsubsection{Classification by Size, Range, and Endurance}

Military UAVs are commonly classified in a unified manner according to their size, operational range, and endurance, as these attributes jointly determine their mission roles and capabilities~\cite{7995044}.

\begin{enumerate}[label=\arabic*., leftmargin=*]
    \item \textit{Micro UAVs (very short range and endurance):}
    Extremely small and lightweight platforms, usually under 2 kg, with ranges of only a few kilometers ($\leq$10 km) and endurance of a few minutes up to about 1 hour. They are primarily used for indoor reconnaissance, close-proximity surveillance, and squad-level missions in urban or cluttered environments. 
    \textit{Example}: The Black Hornet Nano weighs about 18 grams and is used by several military forces for short-range reconnaissance and surveillance.

    \item \textit{Mini UAVs (short range and endurance):}
    Slightly larger systems with typical ranges up to $\sim$50 km and endurance of 1--6 hours. They are commonly deployed for localized reconnaissance, border or perimeter monitoring, and environmental sensing, where modest endurance and portability are advantageous. 
    \textit{Example}: The RQ-11 Raven and Wasp AE are used by the U.S. Army and allied forces for tactical reconnaissance missions.

    \item \textit{Tactical UAVs (medium range and endurance):}
    Medium-sized platforms capable of operating over distances up to 150--650 km with endurance between 8 and 48 hours. They are widely used for battlefield reconnaissance, target acquisition, and combat assessment, providing persistent situational awareness to tactical units. 
    \textit{Example}: The RQ-7 Shadow and Bayraktar TB2 are commonly employed for tactical intelligence and combat support operations.

    \item \textit{Strategic UAVs (long range and endurance):}
    The largest and most capable UAVs, with ranges extending beyond 650 km (often several thousand kilometers) and endurance exceeding 24--48 hours. They are designed for high-altitude, long-duration missions such as persistent intelligence gathering, strategic surveillance, and wide-area monitoring. Examples include the Global Hawk and Predator UAVs. 
    \textit{Example}: The RQ-4 Global Hawk and MQ-9 Reaper (Predator B) are used for long-range strategic intelligence, surveillance, reconnaissance, and precision strike operations.
\end{enumerate}

From a military perspective, these classes highlight critical trade-offs between persistence, survivability, and operational risk. Strategic UAVs offer broad coverage and endurance but are vulnerable in heavily defended environments, while tactical UAVs provide balanced capabilities for battlefield support. Mini UAVs add flexibility at lower logistical cost, and micro UAVs contribute resilience through expendability and swarming tactics, particularly effective in frontline and urban missions. In practice, militaries combine all four classes to achieve both strategic situational awareness and adaptable, redundant support in contested battlespaces.

\vspace{-0.5em}
\subsubsection{Classification by Function} Military UAVs are also classified by their primary function, which includes surveillance, combat, reconnaissance, and logistics. Surveillance UAVs, such as the MQ-1 Predator and RQ-4 Global Hawk, are equipped with advanced cameras and sensors for continuous monitoring and intelligence collection over large areas. Combat UAVs, such as the MQ-9 Reaper and Bayraktar TB2, are armed platforms designed for precision strikes and close air support missions. Reconnaissance UAVs, including the RQ-7 Shadow and RQ-170 Sentinel, are used to gather intelligence on enemy positions, movements, and activities, often operating in contested or hostile environments. Logistics UAVs, such as the Aeralis T-600 and Elroy Air Chaparral, are designed to transport supplies, equipment, and medical aid to remote or high-risk locations where traditional transport methods are impractical or dangerous.

\begin{figure}[t]
    \centering
    \includegraphics[width= 1\linewidth]{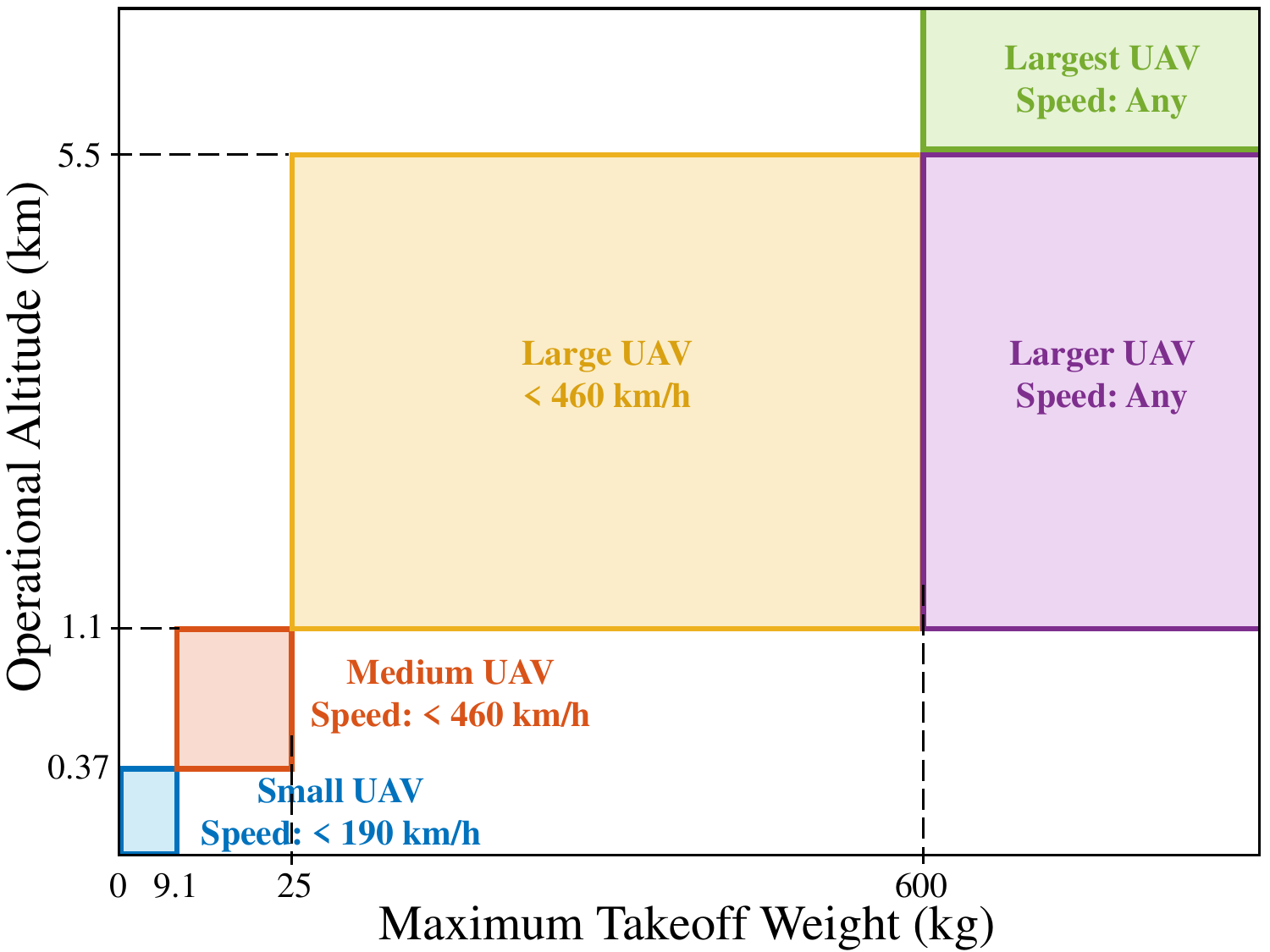} 
    \caption{The U.S. Department of Defense UAV Classifications} 
    \label{fig:DoD_UAV_Classifications}
    \vspace{-2em}
\end{figure} 

\begin{figure}[t]
\centering
\begin{subfigure}[t]{1\linewidth}
    \centering
    \includegraphics[width= 1\linewidth]{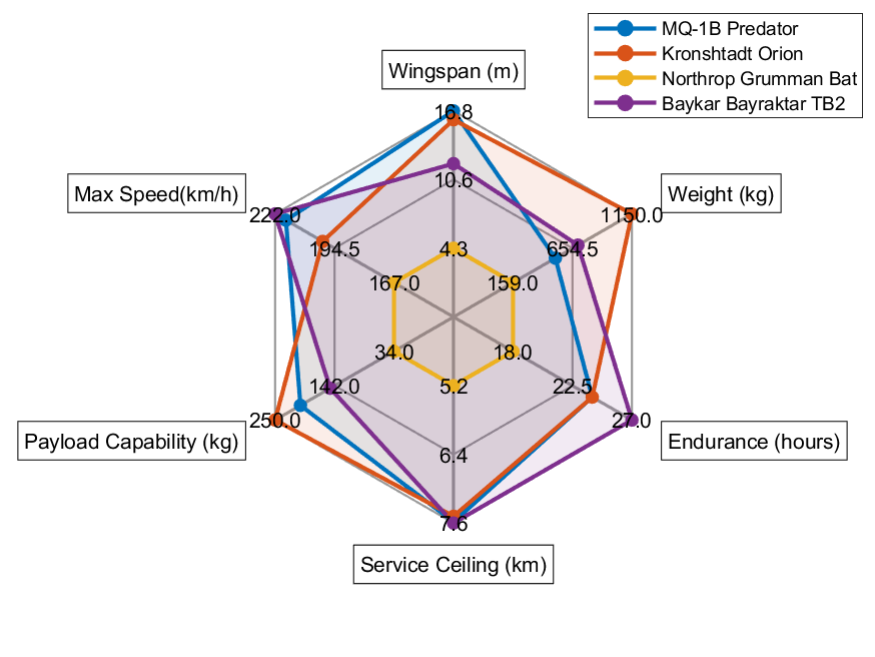} 
    \vspace{-2em}
    \caption{Features of some medium-altitude long-endurance (MALE) UAVs.} 
    \label{sfig:UAV_features_MALE}
\end{subfigure}
\\
\begin{subfigure}[t]{1\linewidth}
    \centering
    \includegraphics[width= 1\linewidth]{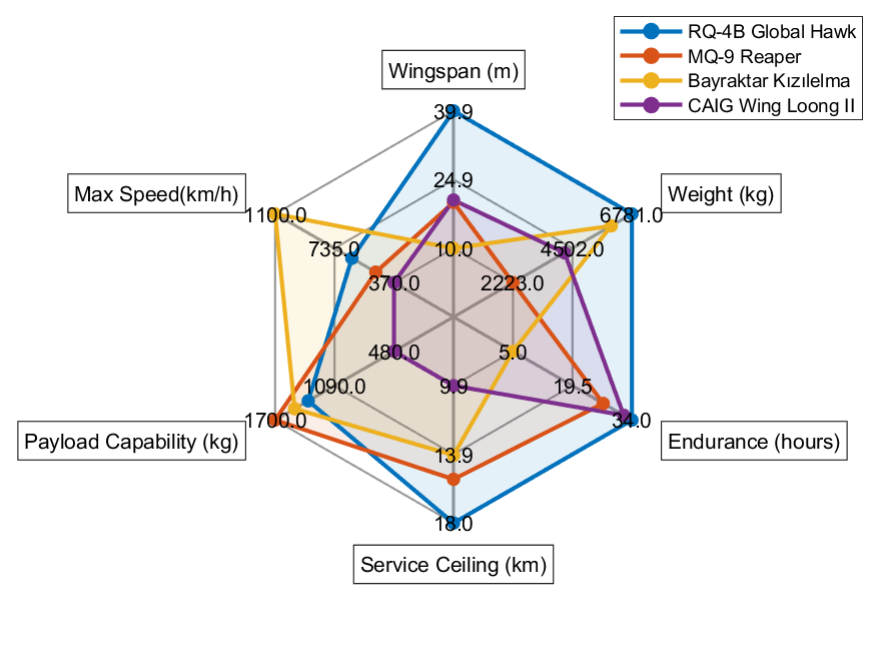}
    \vspace{-2em}
    \caption{Features of some high-altitude long-endurance (HALE) UAVs.} 
    \label{sfig:UAV_features_HALE}
\end{subfigure}
\caption{Features of some UAVs categorized as MALE UAVs whose altitude $\le 9$~km and range$>200$~km, and HALE UAVs whose altitude $>9$~km and having indefinite range.
}
\label{fig:UAV_features_expl}
\vspace{-1em}
\end{figure}

\subsubsection{Classification by Propulsion System} 
Another way to classify military UAVs is by their propulsion system, which includes fixed-wing, rotary-wing, and hybrid types. Fixed-wing UAVs, similar to traditional aircraft, are highly efficient for long-distance and high-altitude missions, providing stable platforms for strategic surveillance, persistent intelligence, surveillance, and reconnaissance (ISR), or wide-area communication relays. Examples include the RQ-4 Global Hawk and MQ-1 Predator, both designed for extended endurance and large-area intelligence missions. Rotary-wing UAVs, such as multirotors and helicopters, offer vertical takeoff and landing capability and sustained hovering, which are essential in confined or urban battlefields where rapid deployment and precise positioning are required. Examples include the MQ-8 Fire Scout and the RQ-16 T-Hawk, which provide close-range surveillance, reconnaissance, and target acquisition support for ground forces. Hybrid UAVs combine both modes, delivering the endurance of fixed-wing designs with the maneuverability of rotary-wing platforms, making them versatile assets for diverse military missions including contested resupply, tactical communications, and EW support. Examples include the V-BAT and the Arcturus Jump 20, both capable of vertical takeoff and efficient cruise flight for extended-range operations.

While propulsion is not directly part of the communication subsystem, it strongly influences performance in military operations. Propulsion demands dominate energy consumption—several orders of magnitude higher than communication requirements \cite{9690481}—and thus directly constrain mission endurance and relay availability. Power limitations may reduce the range, stability, or reliability of communication relays, particularly in extended or high-tempo operations. Furthermore, propulsion-induced vibrations, acoustic signatures, and thermal emissions can affect stealth, sensor payload stability, and detectability, which are critical in contested environments. For these reasons, propulsion choice must be aligned with mission priorities, balancing endurance, mobility, and survivability to ensure reliable communication support in hostile battlespaces.

\begin{figure*}[t]
    \centering
    \includegraphics[width=0.8\textwidth]{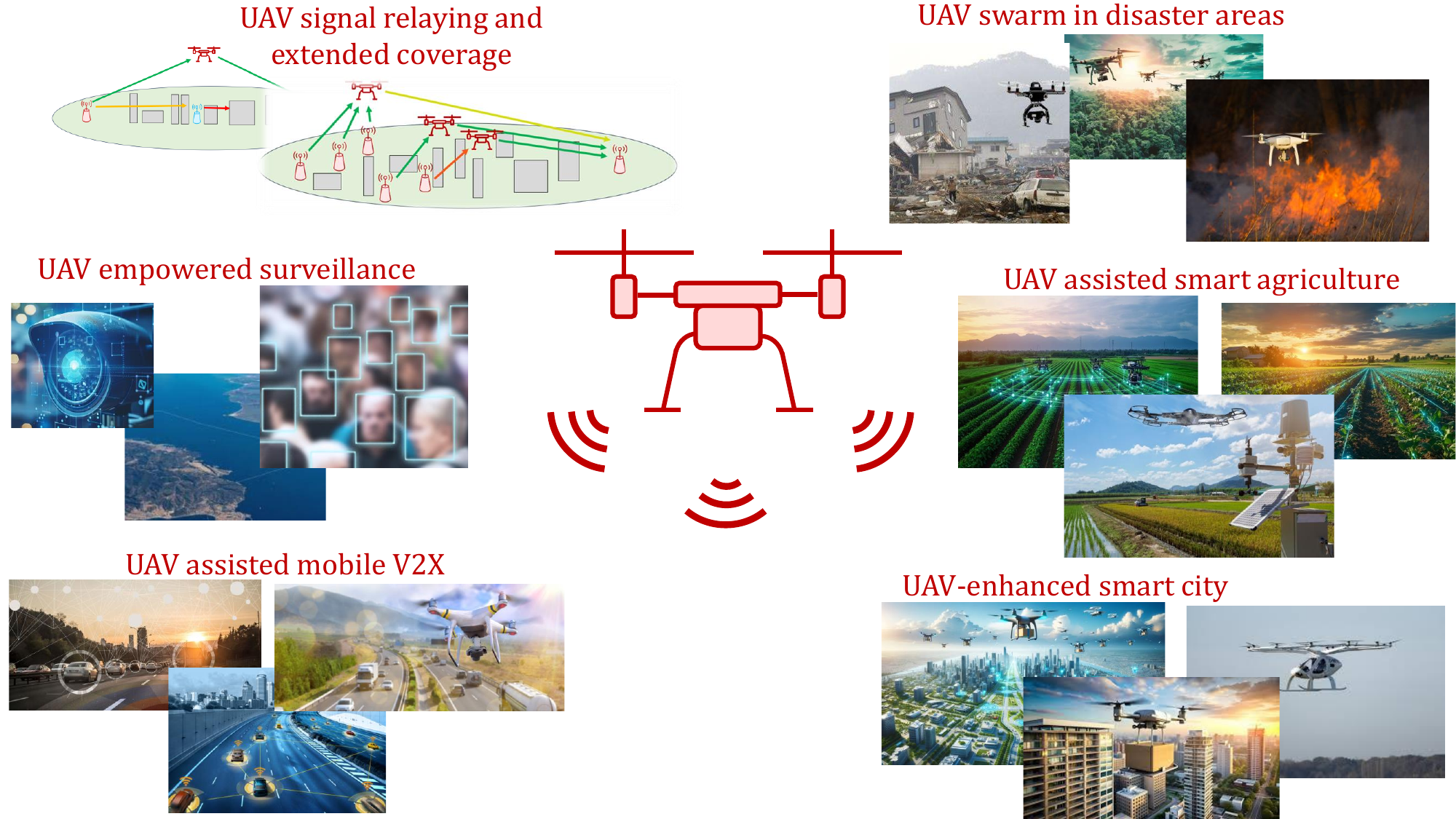}%
    \caption{UAV-assisted wireless networks.} 
    \label{fig:wireless_net}
    \vspace{-1em}
\end{figure*}

\subsubsection{Classification According to the Payload Capacity} In this subsection, Military UAVs are further classified according to their payload capacity as follows \cite{Crouch2005, Fahlstrom2022}:
\begin{enumerate}[label=\alph*)]
    \item \textit{Very Small UAVs:} The very small UAV class includes UAVs ranging from insect-sized to 30-50 cm long. Micro designs with flapping or rotary wings are popular for their lightweight and maneuverability, suitable for spying and biological warfare. Examples include U.S. Aurora Flight Sciences Skate (60 cm wingspan, 33 cm length), and Australian CyberQuad Mini (42x42 cm).
    \item \textit{Small UAVs:} Can carry payloads of 2 kg to 10 kg. These UAVs are often used for more demanding tasks, including precision agriculture, infrastructure inspection, and search and rescue operations. For example, the 1-meter long RQ-11 Raven by AeroVironment, with a wingspan of 1.4 meters, and the RQ-7 Shadow used by the U.S. Army.
    \item \textit{Medium UAVs:} Feature payload capacities around 100-200 kg. They are used in applications requiring substantial equipment, such as larger surveillance systems, delivery services, and advanced scientific research. Examples include the UK Watchkeeper, the RQ-2 Pioneer, and the American Aerospace RS-20.
    \item \textit{Large UAVs:} Have a payload capacity exceeding 200~kg. These UAVs are employed for heavy-duty tasks, including military operations, extensive logistical support, and large-scale industrial inspections. Examples include the US General Atomics Predator A and B and the US Northrop Grumman Global.
    \item \textit{Very Large UAVs:} They are often used in military and commercial sectors for significant logistics operations, aerial refueling, and large equipment transport.
\end{enumerate}

\vspace{-1em}

In military applications, UAV payload classes represent a trade-off between operational capability and risk. Very small and small UAVs carry limited payloads but offer advantages in stealth, expendability, and resilience through swarming, making them well-suited for reconnaissance and tactical relay missions. Medium UAVs can accommodate heavier ISR or EW payloads, but they impose greater logistical demands and are more easily detected. Large and very large UAVs support strategic missions, including long-range ISR, strike operations, and logistical support; however, their size and operational value increase their vulnerability in contested or defended airspace.

\subsubsection{ U.S. Department of Defense classification}
The U.S. Department of Defense classifies UAVs into five categories, as shown in Fig.~\ref{fig:DoD_UAV_Classifications}~\cite{USArmy2010, 9369901}. This classification, based on weight, altitude, and speed, is not merely administrative; it directly informs the relay capability of a platform. For instance, Group 1 UAVs like the RQ-11B Raven offer rapid, tactical-level relay for small units, while Group 2 systems such as the ScanEagle provide enhanced endurance for company-level operations. Group 3 platforms, including the RQ-7B Shadow, offer greater payload for robust battalion-level communications. Groups 4 and 5, represented by the MQ-1C Gray Eagle and RQ-4 Global Hawk respectively, deliver the endurance, power, and sophisticated hardware necessary for persistent, strategic-level communications over vast areas. This hierarchy underscores that UAV relay selection is fundamentally constrained by these doctrinal categories, with each group fulfilling a distinct operational niche.  

\begin{table*}[htbp]
\centering
\caption{Representative Military UAV Platforms and Their Operational Specifications}
\label{tab:uav_applications}
\begin{tabular}{@{}p{3.5cm}p{3.5cm}p{2cm}p{2cm}p{5cm}@{}}
\toprule
\textbf{UAV Platform} & \textbf{Classification} & \textbf{Range} & \textbf{Endurance} & \textbf{Military Applications} \\ \midrule
Black Hornet Nano & Micro (Very Short) & $\leq$ 2 km & $\sim$25 min & Close-proximity Surveillance \\
RQ-11 Raven & Mini (Short) & $\sim$10 km & 60--90 min & Tactical Reconnaissance \\
RQ-20 Puma AE & Mini (Short) & $\sim$15 km & 2--3.5 h & Reconnaissance, SAR \\
ScanEagle & Mini/Tactical & $\sim$100 km & 24+ h & Persistent Surveillance, Tactical Relay \\
RQ-7 Shadow & Tactical (Medium) & $\sim$125 km & 6--8 h & Tactical ISR, Reconnaissance, Relay \\
Bayraktar TB2 & Tactical (Medium) & $\sim$150 km & 27 h & Armed Reconnaissance, Precision Strikes \\
V-BAT & Tactical (Medium) & $\sim$130 km & 10 h & Reconnaissance, Tactical Comms \\
Arcturus Jump 20 & Tactical (Medium) & $\sim$185 km & 13+ h & ISR, Tactical Relay, EW \\
MQ-8 Fire Scout & Tactical (Medium) & $\sim$200 km & 8--12 h & Maritime ISR, Targeting, SAR \\
MQ-1 Predator & Strategic (Long) & $\sim$650 km & 24+ h & Surveillance, Reconnaissance, Strike \\
MQ-9 Reaper & Strategic (Long) & $\sim$1,850 km & 27+ h & Precision Strikes, ISR, Relay \\
RQ-4 Global Hawk & Strategic (Long) & $\sim$22,800 km & 30+ h & Strategic ISR, High-Altitude Relay \\
X-47B & Strategic (Long) & $\sim$3,900 km & 6+ h & UCAV, Stealth Technology Demonstrator \\
 \bottomrule
\end{tabular}
\end{table*}

Similarly, a well-known classification criterion for large UAVs is based on their altitude and endurance, where they are categorized as either medium-altitude long-endurance (MALE) UAVs or high-altitude long-endurance (HALE) UAVs \cite{Watts2012}. Figs.~\ref{fig:UAV_features_expl}(\subref{sfig:UAV_features_MALE}) and~\ref{fig:UAV_features_expl}(\subref{sfig:UAV_features_HALE}) illustrate typical MALE and HALE UAVs along with their distinguishing features. The main criterion for classification is the supported altitude where UAVs with supported altitude less than or equal to 9~km are classified as MALE, otherwise they are classified as HALE. Some sources even use higher altitudes, such as 18~km, to classify UAV as HALE. In such a case, only few UAVs such as the RQ-4[B] Global Hawk is classified as HALE. Furthermore, solar powered, lightweight prototypes/airships/ballons/satellites can be classified as HALE. In general, HALE UAVs can fly at high altitudes and support a long endurance.

\subsection{Military Applications of UAVs} 
     
\begin{figure}[t]
    \centering
    \includegraphics[width=\columnwidth]{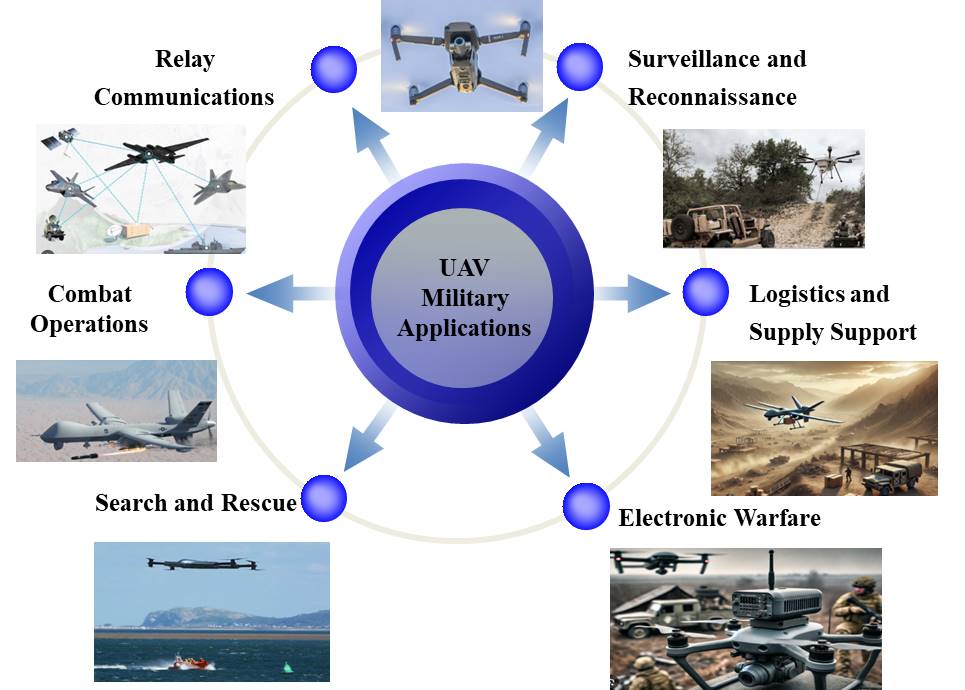} 
    \caption{UAV Military applications.} 
    \label{fig:UAV Military applications}
    \vspace{-1em}
\end{figure}

While UAVs are widely deployed in civilian domains such as disaster response, environmental monitoring, wildfire management, smart-city surveillance, and entertainment \cite{Ruetten2020, CNN2011, 9685991, 8453584, Qadir2021, Saffre2022, Alsammak2022, Kharchenko2022, Alawad2023, 9354917, Flytbase2023, 9922828, Intel2016, Coachella2023}, many of these applications, as shown in Fig.~\ref{fig:wireless_net}, have direct military counterparts. For instance, disaster response capabilities parallel battlefield damage assessment, while smart-city monitoring translates into base-security and force-protection missions. However, simply transferring UAV algorithms and designs from civilian applications to defense contexts is inadequate. Military operations are subject to far more stringent constraints, including exposure to hostile interference, deliberate attempts to disrupt or intercept communications, requirements for secure and covert operation, and the need for robustness in highly contested and unpredictable environments. These challenges necessitate specialized platforms, algorithms, and communication protocols explicitly tailored to the operational demands of defense applications. The diverse military UAV platforms that fulfill these demanding roles, along with their key specifications and primary functions, are summarized in Table~\ref{tab:uav_applications}. The main military applications, as shown in Fig.~\ref{fig:UAV Military applications}, are: 

\subsubsection{Surveillance and Reconnaissance} Military UAVs provide continuous real-time intelligence, gathering critical data through high-resolution cameras and advanced sensors. This enhances situational awareness and informs strategic decision-making, offering commanders a detailed view of the battlefield. UAVs can monitor large areas such as borders, detect enemy movements, and identify potential threats without exposing personnel to danger. For instance, platforms like the MQ-9 Reaper and RQ-4 Global Hawk have been extensively used for persistent surveillance and reconnaissance missions. Smaller UAVs such as the ScanEagle and RQ-11 Raven are often deployed for tactical observation at shorter ranges, providing ground troops with immediate intelligence in dynamic combat environments. The U.S.–Mexico border case study was used to assess the effectiveness of such UAV-based surveillance systems~\cite{Ahmadian2022, 10430396}.

\subsubsection{Combat Operations} In modern combat, UAVs play a vital role, providing capabilities that extend beyond traditional manned platforms. They execute precision strikes against high-value targets, deliver close air support to ground forces, and conduct tactical offensives with reduced risk to human pilots. For example, the MQ-9 Reaper and Bayraktar TB2 have been used effectively for precision strike missions in recent conflicts, demonstrating high accuracy and operational efficiency. UAVs are also employed in battlefield interdiction and suppression of enemy air defenses (SEAD), where platforms such as the IAI Harop loitering munition autonomously seeks and destroys radar emitters. In time-sensitive targeting missions, systems like the XQ-58A Valkyrie and Taranis enable rapid, coordinated responses by leveraging autonomous navigation and real-time data sharing. By combining precision, persistence, and reduced operational risk, UAVs significantly enhance the effectiveness and flexibility of combat operations.

\subsubsection{Logistics and Supply Support} In resupply missions, UAVs are instrumental, especially in hostile or hard-to-reach areas. They deliver essential supplies such as ammunition, medical kits, and food, ensuring that troops remain operational even in isolated locations. For instance, the aerial reconfigurable embedded system (ARES) developed by DARPA and the K-MAX unmanned helicopter have demonstrated autonomous cargo delivery in combat zones, reducing the need for vulnerable ground convoys. Similarly, the Skyf Drone and Elroy Air Chaparral are being explored for tactical resupply and medical evacuation operations in contested environments. This capability is crucial for maintaining the momentum of military campaigns and supporting forces in need while minimizing the risk to supply convoys from enemy attacks~\cite{10430396}.

\subsubsection{Search and Rescue}

In military search and rescue missions, UAVs are essential for rapidly reaching hazardous or inaccessible areas to locate isolated or injured personnel. They provide real-time intelligence to recovery teams, enabling faster decision-making and coordination for medical evacuation under fire or in hostile terrain. Examples include the RQ-11 Raven and RQ-20 Puma, which are frequently used for reconnaissance and locating missing soldiers in dangerous zones, as well as the MQ-8 Fire Scout, which can assist in maritime recovery and casualty evacuation operations. Their ability to operate in contested environments, including battlefields with ongoing threats, makes UAVs indispensable for personnel recovery and life-saving operations in modern warfare~\cite{10430396}.  

\begin{table*}[htbp]
\caption{Comparison Between UAV-Based Relay and UAV-Based BS Communications}
\centering
\resizebox{\textwidth}{!}{%
\begin{tabular}{@{}p{5cm} p{6cm} p{6cm}@{}}
\toprule
\textbf{Aspect} & \textbf{UAV-Based Relay} & \textbf{UAV-Based BS} \\ 
\midrule
Mobility & Highly dynamic, frequently repositioned. & Limited or fixed mobility. \\ 
Latency & Higher from multi-hop links. & Lower with direct links. \\ 
Energy Consumption & Lower power, passive function. & Higher power, active transmission. \\ 
Security & Multiple vulnerable points. & Single critical node. \\ 
Deployment & Needs adaptive routing. & Easier integration. \\ 
Cost & Lower, easy to redeploy. & Higher, infrastructure-based. \\ 
Coverage Area & Extends range dynamically. & Covers fixed region. \\ 
Interference Management & Needs coordination. & Can use advanced mitigation. \\ 
Survivability in Military & High stealth and adaptability. & More vulnerable to attack. \\ 
Network Complexity & Complex multi-hop management. & Simpler centralized control. \\ 
Operational Endurance & Longer from low power use. & Shorter from high power use. \\ 
\bottomrule
\end{tabular}
}
\label{tab:UAV_comparison}
\end{table*}

\subsubsection{Electronic Warfare} 
UAVs are increasingly employed as key assets in EW, where they disrupt enemy communications, jam radar systems, and degrade adversarial C2 networks. They can also conduct GPS spoofing, electronic surveillance, and electromagnetic deception to deny the enemy freedom of action in the spectrum. By neutralizing or exploiting these systems, UAVs protect friendly forces, enable freedom of maneuver, and create windows of opportunity for offensive operations. For example, U.S. MALD-J decoy drones \cite{raytheon_mald} have been deployed to jam and confuse enemy radars, while UAVs in the Ukraine conflict have demonstrated the ability to intercept and disrupt adversarial communications \cite{vgi2023dronesEW}. Their contribution to spectrum dominance makes UAVs essential for achieving information superiority on the modern battlefield.

    \subsubsection{UAV-Based Relay Communications} 
    UAV relays are essential to extend coverage in hostile or remote areas, enabling rapid ad-hoc networks for real-time communication in dynamic battlefields~\cite{9864322}. Their mobility ensures robust links under jamming or obstruction while reducing reliance on ground infrastructure, and they support covert missions through secure, low-detection links. However, these advantages come with challenges such as increased latency due to multi-hop routing, the need for sophisticated routing protocols, heightened security risks, and stringent energy management requirements \cite{9795858}.

Relay communications in military networks can rely on terrestrial relays, such as fixed ground stations, or on aerial relays mounted on UAVs. Terrestrial relays are limited by their static deployment, with performance tied to source–relay–destination geometry, and they remain exposed in contested environments. UAV-mounted relays overcome these constraints through mobility and altitude control, enabling adaptive trajectory design, path-loss compensation, and sustained links in denied areas~\cite{8867956, 8626132}. Their maneuverability further supports interference avoidance, rapid reconfiguration, and evasion of adversarial jamming or targeting~\cite{8941129, 9027102, 9056797, 9140376, 9154440}. 

Practical examples include the RQ-4 Global Hawk and MQ-9 Reaper, which can serve as high-altitude communication relays for extending C2 links, and the RQ-7 Shadow and ScanEagle, which can provide tactical-level relay support for ground units in complex or obstructed terrains. These systems illustrate the operational versatility of UAV-mounted relays in maintaining resilient communications during military operations.

~\\

\subsubsection{UAV-Based Base Station Communications}

In addition to relaying, UAVs can be employed as base stations, operating as airborne mobile cells that provide direct connectivity to ground users \cite{7918510}. Equipped with advanced antenna arrays, multiple RF chains, and high-capacity processing units, they can support multiple simultaneous connections and deliver stable, high-throughput service. Their relatively fixed or predictable positioning ensures consistent coverage and reduced latency compared to relay-based systems. However, in military contexts these platforms face critical challenges: high energy consumption due to complex hardware, vulnerability as high-value communication nodes that adversaries may deliberately target, and difficulties integrating with existing tactical, ground, and satellite networks. These limitations result in increased deployment risks, higher logistical burden, and greater maintenance demands in contested operational environments. 

Table~\ref{tab:UAV_comparison} presents a comparison between the features of UAV-based relays and base stations.

\vspace{-1em}
\subsubsection{GPS in Military UAV Operations} 
GPS is a cornerstone of military UAV operations, providing precise navigation, autonomous flight control, and stable positioning essential for mission success. It enables UAVs to follow pre-defined tactical routes, conduct reconnaissance with high spatial accuracy, and maintain stability during surveillance, targeting, or strike missions~\cite{9615200}. 

In combat search and rescue, GPS allows UAVs to rapidly locate and reach designated coordinates, expediting personnel recovery under hostile conditions. In swarm or multi-UAV deployments, GPS ensures tight synchronization and coordinated maneuvers, which are critical for maintaining formation in contested environments. Overall, GPS enhances precision, autonomy, and operational effectiveness across a wide spectrum of military missions, though its reliance also creates vulnerabilities that adversaries exploit through jamming and spoofing.

\begin{figure}[t]
    \centering
    \includegraphics[width=0.4\textwidth]{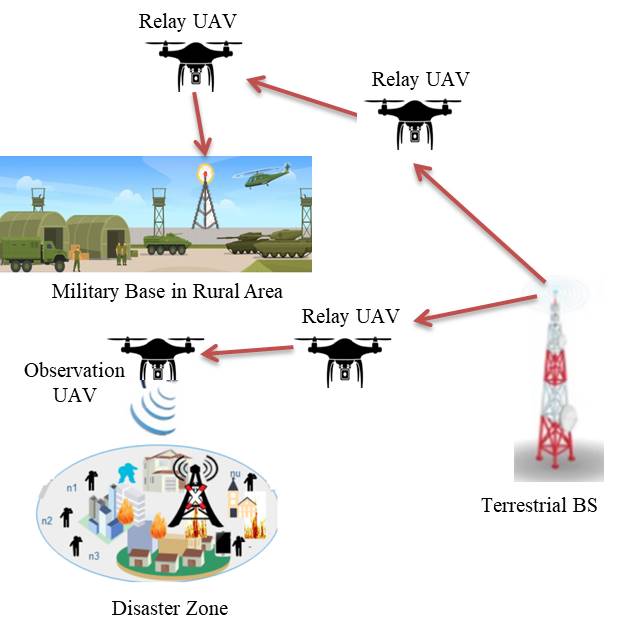} 
    \caption{UAV as an aerial relay.} 
    \label{fig:UAV_as_an_aerial_active_relay}
    \vspace{-1em}
\end{figure}

\begin{table*}[htbp]
\caption{Comparison Between Single UAV and UAV Swarms in Military Operations}
\centering
\resizebox{\textwidth}{!}{%
\begin{tabular}{@{}p{5cm} p{7cm} p{7cm}@{}}
\toprule
\textbf{Aspect} & \textbf{Single UAV} & \textbf{Swarm UAVs} \\ 
\midrule
Control and Coordination & Simple, operator-controlled. & Needs complex, decentralized coordination. \\ 
Task Efficiency & Suited for small, focused missions. & Handles large, multi-objective missions. \\ 
Redundancy and Fault Tolerance & Single failure ends mission. & Redundant; adapts to losses. \\ 
Communication and Information Sharing & Relies on base station. & Peer-to-peer sharing and fusion. \\ 
Survivability in Contested Environments & High risk of detection/elimination. & Dispersed, deceptive, more survivable. \\ 
EW Resilience & Vulnerable to jamming/spoofing. & Resilient via cooperation and agility. \\ 
Cost and Resource Requirements & Lower cost, limited scalability. & Higher cost, efficient for persistent ops. \\ 
Autonomy and Intelligence & Low autonomy, human-in-loop. & High autonomy with ML/adaptive control. \\ 
Mission Complexity & Suited for single-task ops. & Enables complex, multi-domain ops. \\ 
Energy and Payload Capacity & Limited by one UAV. & Shared load across swarm. \\ 
Flexibility and Adaptability & Low adaptability. & Highly adaptable to threats. \\ 
Scalability & Hard to expand without more UAVs/operators. & Easily scalable by swarm size. \\ 
Applications & Recon, strike, comms relay. & ISR, EW, saturation strikes, resilient comms. \\ 
\bottomrule
\end{tabular}
}
\label{tab:comparison_uavs}
\end{table*}

\subsection{Military UAV Regulations} 

Military UAVs operate under defense-specific regulations that prioritize national security, classified missions, and coordination with manned aviation. Oversight is provided by defense ministries and frameworks such as NATO’s STANAG standards \cite{stanag4586}, which ensure interoperability and safety across allied forces. These regulations emphasize airspace deconfliction with civilian aviation and establish rules of engagement, spectrum allocation for secure communications, and compliance with the law of armed conflict. 

In contrast, civilian regulations are fundamental to ensuring safe, secure, and responsible drone operations, addressing airspace safety, collision avoidance, and ethical considerations. Core elements include applicability (UAV class and mission role), operational limits (altitude ceilings, restricted zones, and flight permissions), administrative procedures (licensing, registration, certification), and technical standards (safety, navigation, communication) \cite{Fotouhi2018, Stocker2017}. In the United States, oversight is primarily managed by the FAA and NASA, while in Canada, Transport Canada mandates registration for drones between 250 g and 25 kg and classifies operations as basic or advanced based on risk \cite{TransportCanada2024}.

\subsection{Military UAV Standardization}
Standardization is essential to ensure the interoperability of UAV systems in joint and coalition military operations. NATO STANAG 4586 establishes a unified framework for UAV control. It defines common architectures, interfaces, and communication protocols, enabling integration across heterogeneous UAVs and ground control stations \cite{stanag4586}. Complementary initiatives, such as the TSDSI STD 5002, introduce scalable relay architectures that enhance multi-hop networking, UAV mobility, and airborne relay functions, ensuring adaptability in contested and rapidly evolving operational environments \cite{TSDSI5002}. In parallel, the 3GPP has advanced UAV-specific provisions through Release 17 (TS 22.125), which addresses airspace integration, identification, tracking, and secure communications for UAS operations, while Release 18 (TR 21.918) builds on this foundation by refining mobility, interoperability, and network resilience features essential for mission-critical deployments \cite{3gppTS22125,3gppTR21918}. Collectively, these standardization efforts provide a robust and harmonized foundation for interoperable, resilient, and mission-ready UAV systems capable of supporting a wide range of defense applications.

\subsection{Military-Specific Operational Considerations}
This subsection addresses the unique operational environments and constraints that define military UAV deployments, distinguishing them from civilian applications. It highlights the necessity for specialized design and operational considerations when employing UAV-mounted relays in contested and hostile battlespaces.
\subsubsection{Anti-Access/Area Denial (A2/AD) Environments} 
Anti-Access/Area-Denial (A2/AD) environments use layered defenses—such as long-range missiles, EW, and persistent ISR—to limit an adversary’s ability to enter or operate in a region. The goal is to increase risk and deter intervention rather than win through direct engagement. In such contested areas, UAVs play important roles on both sides. For attacking forces, they provide low-risk communication relays that help maintain C2 and situational awareness despite jamming or damaged infrastructure. For defending forces, UAVs support ISR, tracking, and deception within the A2/AD network. As these systems expand, militaries increasingly rely on flexible UAV deployments and manned–unmanned teaming to operate effectively in highly contested airspace \cite{Schmidt2017A2AD}.

\begin{table*}[t]
\centering
\caption{Research Papers on UAVs – Technical Focus and Military Deployment Gaps}
\label{tab:research_papers_gaps}
\footnotesize
\begin{tabular}{p{5cm} p{5.5cm} p{6cm}}
\toprule
\textbf{Research Papers} & \textbf{Technical Focus} & \textbf{Military Gap} \\
\midrule
\cite{8758183}, \cite{9358097}, \cite{8867956}, \cite{7918510}, \cite{9149403}, \cite{8903530}, \cite{10122731}, \cite{10478734}
& UAV Placement \& Coverage Optimization 
& Lacks threat-aware positioning; neglects stealth and EW considerations. \\
\midrule
\cite{8959174}, \cite{9027102}, \cite{7875081}, \cite{8641313}, \cite{9767553}, \cite{9454372}, \cite{10612249}
& Trajectory \& Mobility Optimization
& Does not model evasion under jamming or threat-adaptive path planning. \\
\midrule
\cite{7317490}, \cite{8709739}, \cite{zhao2021outage}, \cite{9599478}, \cite{10449453}, \cite{10449453}
& Channel Modeling (A2G, RIS-assisted) 
& Omits battlefield effects: EMI, clutter, and dynamic jamming environments. \\
\midrule
\cite{8501974}, \cite{9034071}, \cite{8611204}, \cite{10146504}, \cite{10283889}, \cite{10070838}, \cite{9538830}, \cite{10214219}
& Physical Layer Security \& Secrecy 
& Focuses on passive eavesdropping, not active jamming or LPI/LPD in contested spectra. \\
\midrule
\cite{9864322}, \cite{8626132}, \cite{9154440}, \cite{armytech2025LPD}, \cite{10586977},  \cite{khalil2023blos}, \cite{9902968}
& Covert Comms \& LPI/LPD 
& Not integrated with adaptive spectrum tactics for mission-aware stealth. \\
\midrule
\cite{yao2023}, \cite{9989438}, \cite{10289638}, \cite{9386233}, \cite{9778241}, \cite{10462233}, \cite{8756296}, \cite{8941129}
& Energy Efficiency \& RIS-UAV Integration 
& Does not address battlefield power/survival trade-offs or RIS performance under jamming. \\
\midrule
\cite{10757329},\cite{10283826},\cite{9767553}, \cite{9826431},   \cite{8255739},\cite{10045049}, \cite{10190734},  \cite{10371218}
& Machine Learning for UAV Control 
& Lacks robustness to adversarial EW and real-time adaptation in degraded environments. \\
\midrule
\cite{10430396}, \cite{9777886}, \cite{10668867}, \cite{9834117}, \cite{10045049}, \cite{10854532}, \cite{9795858}, \cite{9826431}
& UAV Swarm \& Multi-UAV Coordination 
& Misses resilience to attrition and decentralized anti-jamming in contested networks. \\
\midrule
\cite{khan2021role}, \cite{9056797}, \cite{9140376}, \cite{9778241}, \cite{10533208}, \cite{8708975}, \cite{10345491}
& Interference Management \& Resource Allocation 
& Assumes cooperative interference; ignores tactical QoS under hostile electronic attack. \\
\bottomrule
\end{tabular}
\end{table*}

\subsubsection{Integration with Tactical Data Links (TDLs)} 
A key element enabling the effective use of UAVs in A2/AD environments is their seamless integration with Tactical Data Links (TDLs). TDLs are standardized communication systems (such as Link 16 or Link 22) that allow different military platforms to share real-time situational information, target data, and command and control (C2) instructions digitally and securely. In an A2/AD scenario, where adversaries actively employ EW to disrupt communications, TDL integration provides several critical advantages: data gathered by a single UAV acting as a forward sensor is immediately available to all networked assets, accelerating the targeting cycle and enhancing the overall common operational picture. TDLs are designed with features like frequency hopping and robust encryption, which provide inherent resilience against jamming attempts typical of A2/AD strategies. Furthermore, UAVs can function as elevated, mobile TDL relays, extending the network's range and integrity beyond the usual line-of-sight limitations of many tactical radios, pushing critical data deeper into contested territory. By using secure, machine-to-machine data exchanges, forces can minimize reliance on more easily intercepted voice communications, thereby enhancing operational security within high-threat areas \cite{Schmidt2017A2AD}.

\subsection{Channel Modeling Aspects for Military UAVs}
Reliable UAV relay performance in military operations depends on accurate modeling of three tightly coupled elements: Air-to-Ground (A2G) propagation, background interference, and intentional jamming. Unlike commercial or civilian settings, military environments introduce terrain masking, mobility, electronic attack, and A2/AD threats that fundamentally change channel behavior. As a result, military-grade channel models must account for blocked LoS conditions, adversarial interference, dynamic battlespace geometry, and platform maneuverability. The following subsections summarize the key modeling considerations required to capture these effects in military UAV  networks.

\subsubsection{Air-to-Ground (A2G) Channel Characteristics}

Air-to-Ground links are central to UAV relay operations, and their behavior differs significantly from terrestrial channels due to elevated UAV positions, variable elevation angles, and highly dynamic geometries. Military A2G modeling must account for terrain masking, platform mobility, and adversarial interference—factors often absent in commercial UAV-cellular studies. Although medium and high altitudes yield a high probability of LoS communication, cluttered urban areas, forests, mountains, and A2/AD environments can greatly increase NLoS conditions and shadowing. Existing A2G literature classifies propagation models into deterministic, stochastic, and geometry-based stochastic approaches. Deterministic ray-tracing offers accurate site-specific predictions; stochastic models characterize average fading such as Rician and log-normal shadowing; and geometry-based stochastic models balance physical realism with tractability. Key operational parameters—including frequency band, elevation-angle-dependent LoS probability, atmospheric attenuation, and UAV motion—must be incorporated to accurately represent dynamic military communication links. These A2G characteristics are critical for designing resilient UAV relays capable of maintaining connectivity in complex and contested battlespaces \cite{8411465,8709739}.

\subsubsection{Interference and Jamming Models}

Military UAVs must operate in dense electromagnetic environments where both unintentional interference and deliberate electronic attack can degrade link reliability. Background interference may originate from friendly emitters, ground forces, airborne assets, and co-located RF systems, and is commonly modeled using stochastic geometry or co-site interference frameworks to capture aggregate effects. Unlike civilian interference models, military scenarios must also consider spectrum congestion from multiple tactical radios, radars, and data links operating simultaneously in fast-changing operational settings.

Jamming models address intentional, adversary-driven disruption. Military jamming assumes an adaptive, goal-driven attacker capable of exploiting UAV vulnerabilities through barrage noise jamming, spot and swept jamming, or deceptive techniques such as GPS spoofing and tactical data-link message injection. These models characterize the jammer’s waveform, power, placement, and strategy, enabling analysis of UAV communication vulnerability and high-risk zones. They also guide the design of countermeasures including frequency hopping, adaptive power control, and antenna nulling techniques required to maintain reliable relay performance in contested airspace \cite{9900257}.

\subsection{Military Gaps in Existing UAV Research}

Although UAV communication research has progressed across areas such as placement optimization, trajectory control, channel modeling, physical-layer security, and swarm coordination, most existing studies are developed under benign, cooperative, or interference-limited assumptions. Military operations, however, impose far harsher conditions—characterized by EW, intentional jamming, spectrum denial, stealth and LPI/LPD constraints, multi-domain coordination, and real-time threat adaptation. As shown in Table~\ref{tab:research_papers_gaps}, current research rarely incorporates these operational realities, resulting in models and algorithms that may perform well in civilian or controlled scenarios but fall short in contested or A2/AD environments. Identifying these gaps is essential for guiding future work toward UAV relay systems that are resilient, threat-aware, and suitable for modern battlefield deployment.

\begin{figure*}[t]
    \centering
    \includegraphics[width=0.7\textwidth]{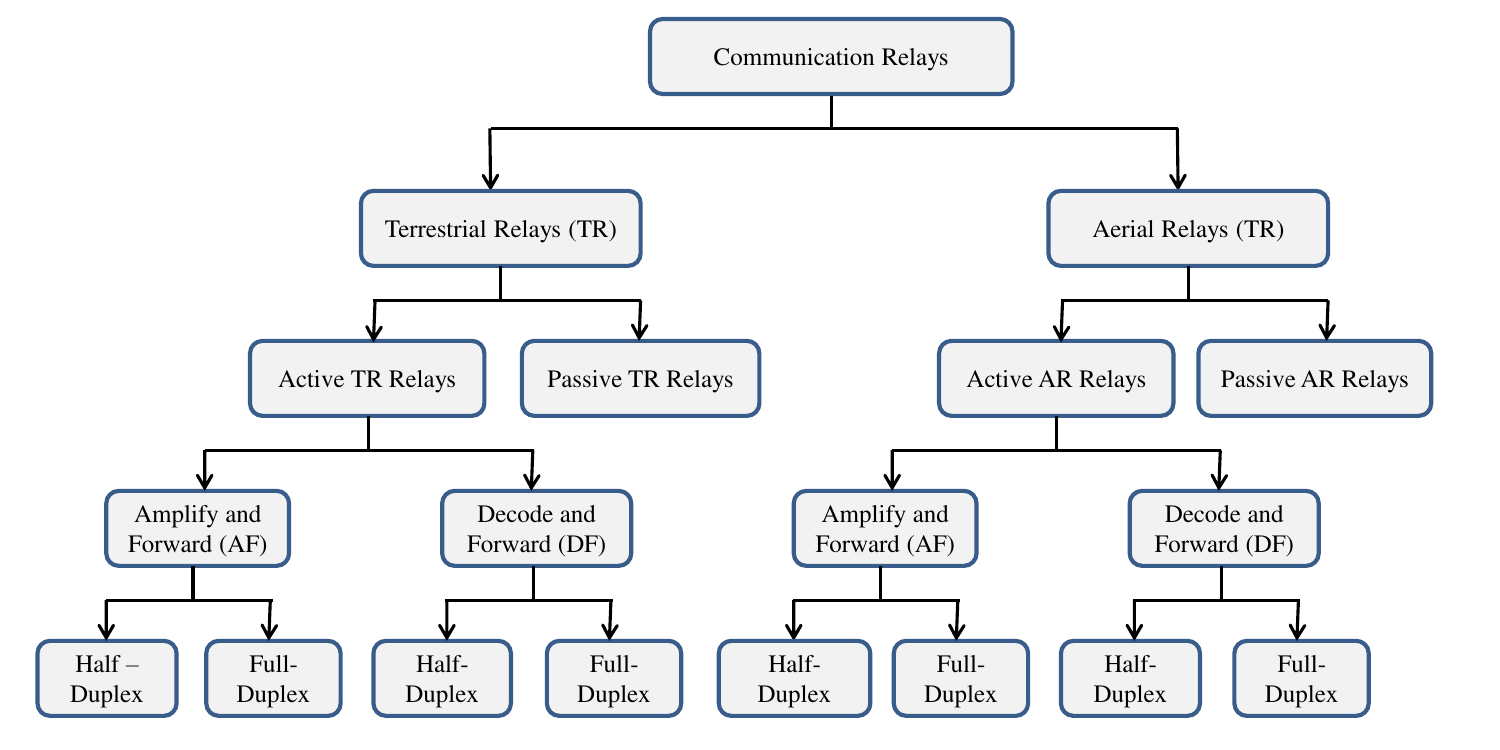} 
    \caption{\textcolor{black}{Classifications of the communication relays.}} 
    \label{fig:Classifications of the communication relays}
    \vspace{-2em}
\end{figure*} 

\subsection{Lessons Learned}
  The evolution and classification of UAVs show that their military effectiveness is determined by the specific mission requirements. Micro and mini UAVs deliver agility and stealth for reconnaissance and close-support tasks but remain limited by payload capacity and survivability against hostile fire. MALE and HALE systems provide long endurance and strategic reach, enabling persistent surveillance and wide-area coverage, but introduce significant logistical demands and increased vulnerability to advanced air defenses. Reliance on single UAV platforms exposes operations to mission failure, underscoring the need for swarm-based concepts that offer redundancy, self-healing communications, and adaptive execution under attrition. UAV-mounted relays and aerial base stations extend the reach and robustness of C2 networks, but their operational impact is constrained by endurance limits, energy requirements, and susceptibility to jamming and electronic attack. The key takeaway is that effective UAV employment demands precise alignment of platform selection with mission objectives—balancing mobility, survivability, endurance, and payload—to preserve superiority and resilience in contested battlespaces.

\vspace{-1em}

\section{Communication Relays}

     
\subsection{Relay Types} 	

\noindent
Relays in military networks serve as intermediate nodes that extend coverage, maintain connectivity, and enhance communication resilience in contested and obstructed environments. They are essential for sustaining C2 links, mitigating path loss, and ensuring network survivability under EW and adversarial conditions~\cite{1362898, Nomikos2017}.

Relays can be classified into the following categories ~\cite{gu2023survey}:

\begin{enumerate}
    
     \item \textbf{Deployment:} Relays can be classified into terrestrial and aerial deployments. TRs, such as AN/TRC-170 and IRR-B4, provide stable coverage and high throughput but are limited by terrain, mobility, and vulnerability to targeting or strikes. ARs, including UAV-mounted platforms (e.g., RQ-7 Shadow UAV or MQ-9 relays), offer rapid redeployment, adaptive altitude control, and the ability to maintain line-of-sight links in obstructed or contested areas.

     \item \textbf{Operational Mode:} Relays can be broadly classified into active and passive types. AARs are extensively employed in military networks to support long-range, high-mobility, and tactical communications, where signal regeneration, error correction, and reliable C2 links are vital under EW conditions. Examples include UAV-based relay systems integrated into platforms like the MQ-9 Reaper and RQ-7 Shadow, which extend communication coverage and maintain connectivity between dispersed units. In contrast, passive relays, such as ARIS, do not amplify or decode the signal but instead manipulate the propagation environment through programmable reflections. This enables energy-efficient, low-signature performance, making ARIS particularly advantageous in covert operations, infrastructure-denied environments, and electromagnetically congested or contested areas.
    
    \item \textbf{Type:} Relays can be classified into AF and DF types. AF relays offer rapid, low-latency forwarding but amplify noise, making them suitable for short-range or time-sensitive links. DF relays improve signal integrity and correct errors, which is critical in contested environments where reliable C2 links must be maintained under electronic attack. The choice between AF and DF reflects a trade-off between speed and robustness, with DF preferred for long-range or multi-hop battlefield communications.
    
    \item \textbf{Duplex Mode:} Relays can be classified into Half-Duplex (HD) and Full-Duplex (FD) modes. HD relays are favored in tactical deployments due to simplicity, lower susceptibility to jamming, and robustness under EW conditions. FD relays provide higher spectral efficiency but require advanced interference cancellation, which can be unreliable in contested scenarios.  
\end{enumerate}

\noindent
From a military perspective, integrating TRs and ARs enables layered communication architectures that enhance survivability, reduce detection risk, and maintain connectivity for dispersed forces. ARs compensate when ground infrastructure is degraded, destroyed, or denied, while TRs anchor network stability in secure rear areas. This hybrid approach is increasingly essential for expeditionary, maneuver, and joint-force operations, where resilience, mobility, and adaptability outweigh efficiency alone. Fig.~\ref{fig:Classifications of the communication relays} shows the classifications of the communication relays.

\subsection{Relay Benefits, Challenges and Applications} 
\textbf{Benefits:} Relays play a pivotal role in military communications by extending coverage, enhancing reliability, and maintaining connectivity across complex terrains and contested environments. They provide additional transmission paths that overcome line-of-sight limitations, ensuring that dispersed units remain connected even in remote, rugged, or jamming-prone areas~\cite{gu2023survey}. 
Aerial UAV relays offer dynamic deployment and mobility advantages. For instance, the OnDrone framework~\cite{8758183} demonstrates that dynamically repositioning UAVs using extremal-optimization and Bézier-curve-based trajectories can improve coverage by up to 47\% compared to static straight-line paths. Similarly, studies show that flying UAVs at optimized altitudes (e.g., 100 m) can increase coverage ratios by over 40\% compared to typical pedestrian heights, highlighting the importance of altitude and placement in mission planning~\cite{9149403}. Even though these findings were obtained in non-military scenarios and military operations present different challenges, adaptive UAV placement and altitude optimization can significantly enhance battlefield communication reliability and coverage. Therefore, these techniques must be adapted to the military context, with modifications to address factors such as jamming, EW, limited situational awareness, and dynamic mission requirements in contested or infrastructure-limited environments.

\textbf{Challenges:}  Active UAV relays require substantial power, sophisticated signal processing, and careful mobility management, which can complicate real-time operations. Terrestrial relays are limited by environmental obstacles and vulnerability to targeting, while aerial relays contend with battery constraints and adverse weather~\cite{mohsan2023}. Moreover, many optimization frameworks assume centralized control and perfect knowledge of user positions, which may not hold in contested military scenarios, limiting scalability and resilience~\cite{8758183}.

\textbf{Applications:} Relays support a range of military applications. They enable beyond-line-of-sight communication for maneuvering forces, resilient data transfer under EW, and high-throughput backhaul for ISR operations. UAV relays, in particular, provide rapid, adaptive coverage in dynamic operational theaters, making them indispensable for expeditionary missions, disaster response, and covert operations~\cite{gu2023survey}.

\subsection{Lessons Learned}

The analysis of communication relays reveal that TRs can provide stable and high-capacity links but lack mobility and are susceptible to targeting, whereas ARs can offer agility and adaptable line-of-sight links at the cost of endurance and energy consumption. Among ARs, AARs excel in maintaining robust and regenerative C2 links under EW but impose high power and processing demands. Conversely, ARIS can provide energy-efficient, low-signature operation suited for covert or infrastructure-denied environments, though their passive nature limits range and processing capability. The key insight is that hybrid integration of TRs, and ARs—balancing endurance, stealth, mobility, and resilience—forms the foundation of next-generation, adaptive, and survivable military communication networks.

\section{Aerial Relays}
    Communication relays, also known as ARs, are essential for improving network coverage and are expected to become more common in the coming years \cite{gu2023survey}. This is because they can greatly improve wireless network coverage and capacity. Their advantages include the ability to change altitude, move easily in three dimensions, low cost, and easy deployment within wireless networks. We categorize ARs into AARs and ARIS . AARs actively process and enhance signals through amplification or decoding, while ARIS, simply redirect signals without processing them. 

\subsection{Active Aerial Relays (AARs)} 

\subsubsection{Definition and Operation}
AAR mounted on UAVs extend communication by not only forwarding but also processing signals in-flight ~\cite{10261452}. Unlike passive surfaces, which merely reflect signals, AAR systems actively amplify or decode them before retransmission. Depending on the mode, they either boost the received waveform or decode and re-encode it to deliver cleaner links. This active operation provides stronger, more reliable connections in complex or contested environments, though at the expense of higher energy and system complexity.

\subsubsection{State of the art}
 
    The use of UAVs as mobile relays has gained significant attention in recent years, particularly for enhancing communication in challenging environments. Various studies have explored different strategies to optimize UAV deployment and improve system performance across a range of applications. In addition to the studies mentioned in the introduction, numerous research works have addressed the usage of UAVs. For example, in~\cite{7562472}, UAVs were used as mobile relays to restore communication links among ground stations in disaster scenarios, enhancing outage probability and information rate. In \cite{8629002}, a UAV served as a DF relay to improve command transmission between a controller and a distant robot, optimizing its location and block length to minimize decoding error probabilities. Furthermore, \cite{8951059} optimized a UAV's two-dimensional position to improve satellite-to-terrestrial communication. 
    
    In~\cite{9057748}, the UAV's three-dimensional placement was designed to minimize high-SNR outage probability. The research in~\cite{9027102} used solar UAVs to enhance ship communication, while~\cite{9056797} deployed UAV relays in multi-cell networks to improve cell-edge connectivity. Research on UAV relays has focused on optimizing placement and deployment to improve system performance, addressing challenges like interference, power consumption, and user fairness. For example, \cite{9140376, 8952664, 8867956, 8903530} have developed algorithms for optimizing UAV positioning and trajectory to enhance SNR, minimize power use, and boost coverage in obstructed areas.
    
    Recently, there has been significant interest in using UAVs with NOMA~\cite{10494323}. Studies such as~\cite{10108290} used UAVs with NOMA in maritime IoT networks for efficient signal reception. The studies referenced in \cite{baek2018optimal} and \cite{8769542} applied NOMA to improve throughput and extend coverage by using UAVs to relay signals between base stations and users. Additionally, \cite{8952664} and \cite{9257576} explored UAVs with NOMA for downlink communications and cell edge performance, respectively. Cooperative NOMA systems with UAVs have been investigated to improve communication quality as shown in \cite{7454773} and \cite{tsipi2023machine}. Finally, \cite{9968298} used machine learning to optimize MIMO NOMA systems with UAVs.

\vspace{-2em}

\subsubsection{\textit{Security for Aerial Relay}}
     UAVs can improve wireless network performance, but they also face higher security risks. However, these risks can be reduced by using UAV's mobility to strengthen connections with authorized users and accurately identify potential eavesdroppers~\cite{gu2023survey}. 
     
     Recent advancements in physical layer security for UAV relaying networks focus on optimizing trajectory design, resource allocation, and cooperation to maximize secrecy capacity and minimize secrecy outage probability. For instance, studies~\cite{8501974, 8884126, 9034071, 7875081} address illegal eavesdropping by exploring power allocation and UAV trajectory design to enhance secrecy capacity. Some approaches use dynamic scheduling and jammer UAVs to disrupt eavesdropping, while others integrate local caching and dynamic UAV positioning to improve security~\cite{8626132, 9154440}. 

The work in \cite{9864322} studies a covert communication scheme employing UAV relays, with a specific focus on hiding range enhancement. However, unlike our comprehensive analysis, their work does not provide a broader review or comparative evaluation of relay architectures.
     
     Additionally, methods like power splitting and adaptive UAV placement help optimize secrecy in various scenarios, including mmWave and hybrid satellite-terrestrial networks~\cite{8611204, 9075988, 8788530, 8641313}. Despite these advancements, challenges remain with passive and active eavesdroppers, and further research is needed to address these issues and enhance security in UAV networks~\cite{8932449, 9122500}.


\begin{figure}[t]
\centering
\begin{subfigure}[t]{0.48\linewidth}
    \centering
    \includegraphics[width= 1\linewidth]{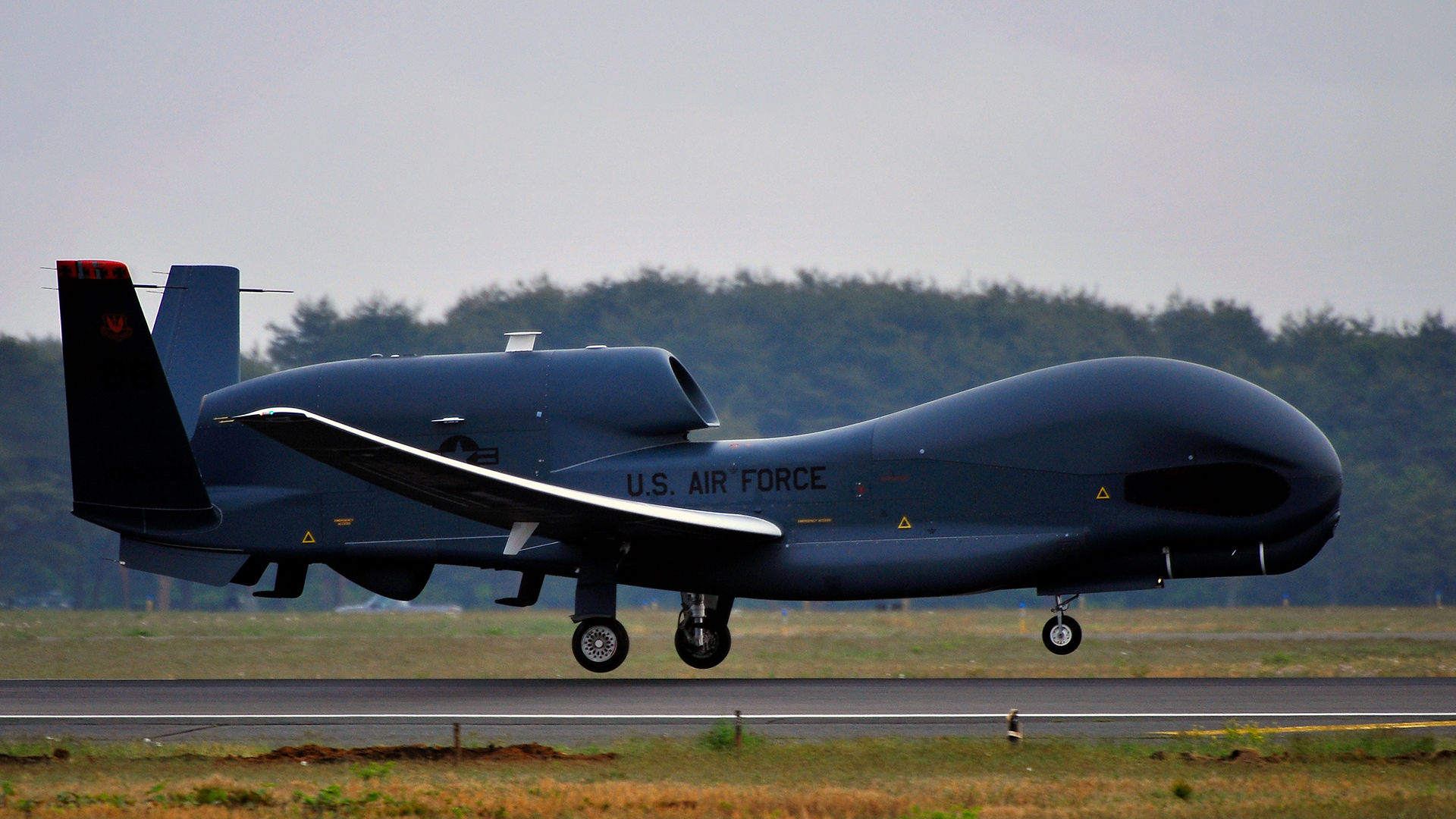} 
    \vspace{-1em}
    \caption{RQ-4 Global Hawk} 
    \label{subfig:RQ4}
\end{subfigure}
\begin{subfigure}[t]{0.48\linewidth}
    \centering
    \includegraphics[width= 1\linewidth]{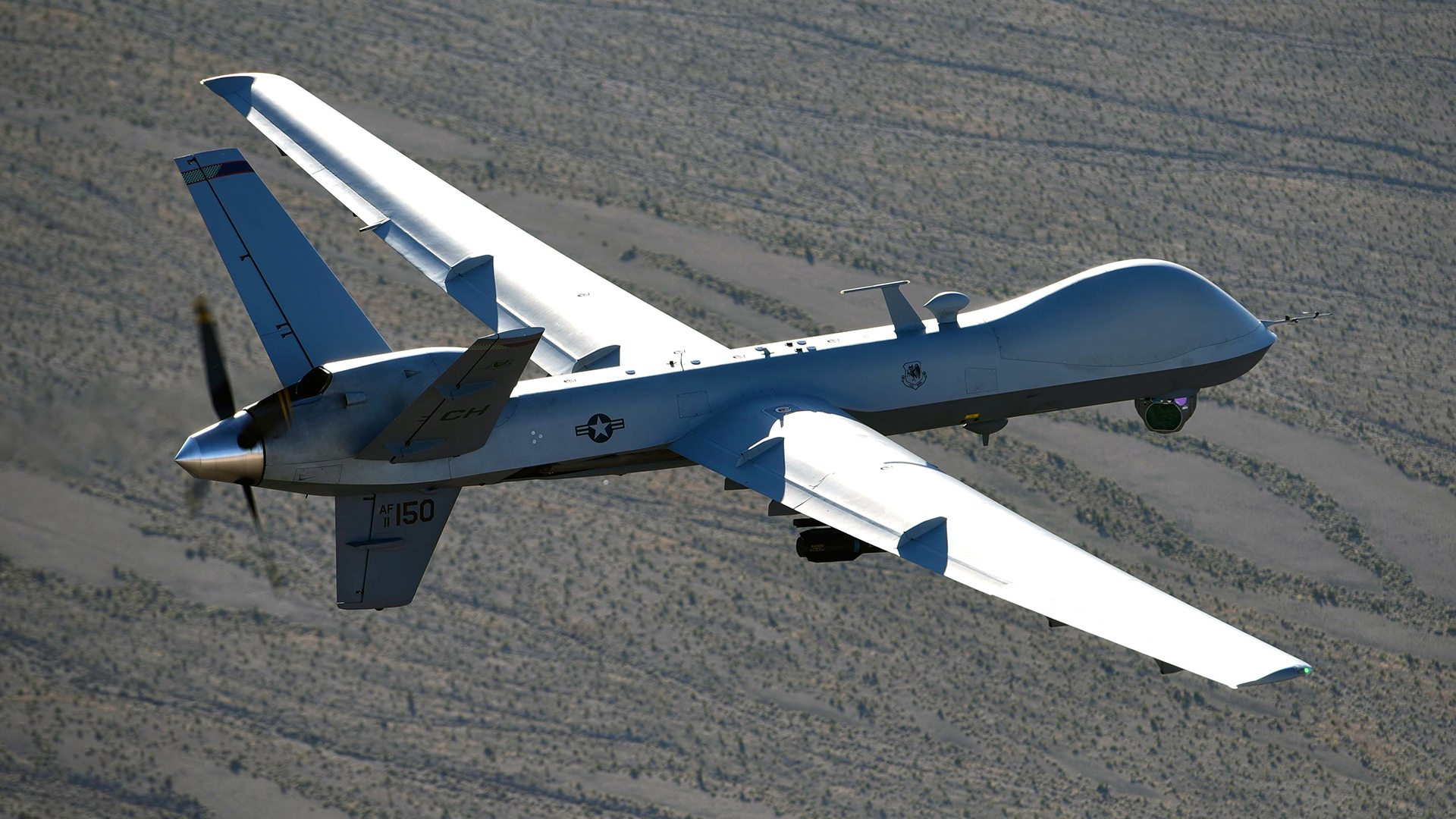}
    \vspace{-1em}
    \caption{MQ-9 Reaper} 
    \label{subfig:MQ9}
\end{subfigure}
\caption{Example of multipurpose MALE$\backslash$ HALE UAVs. Source: {https://www.af.mil}}
\label{fig:UAV_example}
\vspace{-1em}
\end{figure} 

\vspace{-1em}

\subsubsection{Military Applications of Aerial Relays}
UAV-based aerial relays provide critical advantages in modern military operations by extending coverage, enabling interoperability, and ensuring resilient communications in contested environments. Their main applications can be summarized as follows:  
\\
\begin{enumerate}
    \item \textit{Beyond Line-of-Sight Communication:}  
    By operating at higher altitude, UAV relays can maintain secure C2 links across obstructed or remote terrains. This ensures reliable connectivity for ground forces, maritime units, and other UAVs during long-range and coordinated missions \cite{khalil2023blos}.  
    \item \textit{Tactical Data Link  Interoperability:}  
    UAV relays can act as mobile gateways between disparate systems such as Link-16 and Link-22, enabling protocol translation and integration of legacy platforms. This strengthens the common operational picture and enhances coalition coordination.  
    \item \textit{Resilient and Adaptable Networks:}  
    Aerial relays can form mobile ad-hoc networks that automatically reroute traffic when nodes are degraded or destroyed. Their rapid deployability and flexibility provide survivable communications in contested environments where terrestrial infrastructure is unavailable or compromised \cite{10746646}. 
    \item \textit{Coverage Extension:}  
    UAV relays can mitigate dead zones in urban warfare, mountainous regions, or maritime operations by providing elevated links. This capability expands situational awareness and guarantees persistent connectivity in otherwise inaccessible areas  ~\cite{gu2023survey, 10531095}.  
    \item \textit{Enhancing Connectivity in Urban Warfare:} In complex urban environments with dense buildings and potential signal obstructions, aerial relays can enhance communication by providing signal amplification and relay capabilities. This helps maintain robust communication links between units dispersed across different buildings or areas.
    \item \textit{Electronic Warfare Support:}  
    UAV relays can enhance spectrum resilience through anti-jamming measures such as frequency hopping, adaptive beamforming, and cognitive radio. Additionally, they contribute to electronic intelligence by intercepting adversary transmissions and supporting spectrum dominance.  
\end{enumerate}

\begin{figure}[t]
\centering
\begin{subfigure}[t]{1\linewidth}
    \centering
    \includegraphics[width= 1\linewidth]{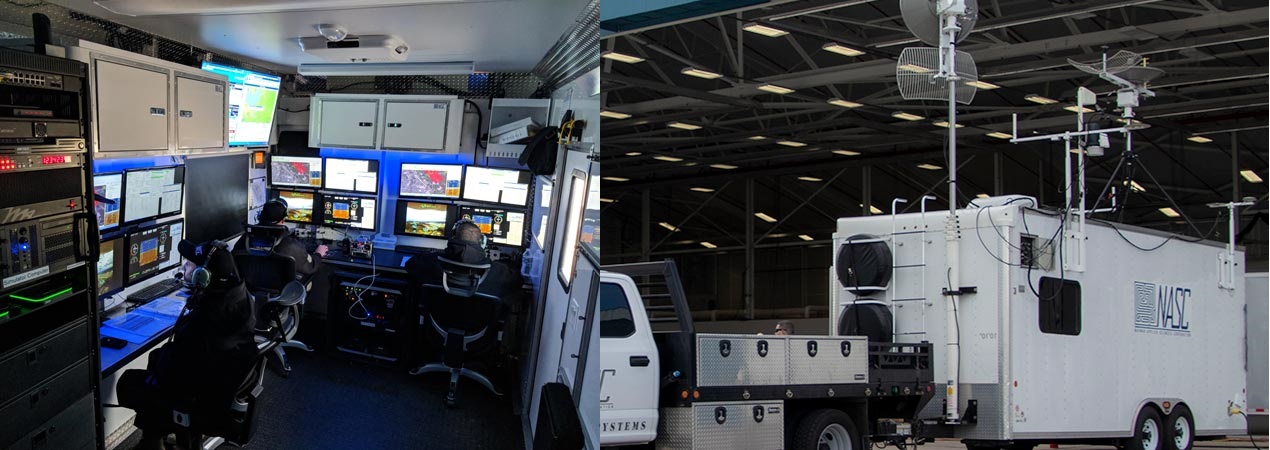} 
    \caption{A mobile operations center for UAV control. Source: Navmar Applied Sciences Corp (NASC)} 
\end{subfigure}
\\
\begin{subfigure}[t]{1\linewidth}
    \centering
    \includegraphics[width= 1\linewidth]{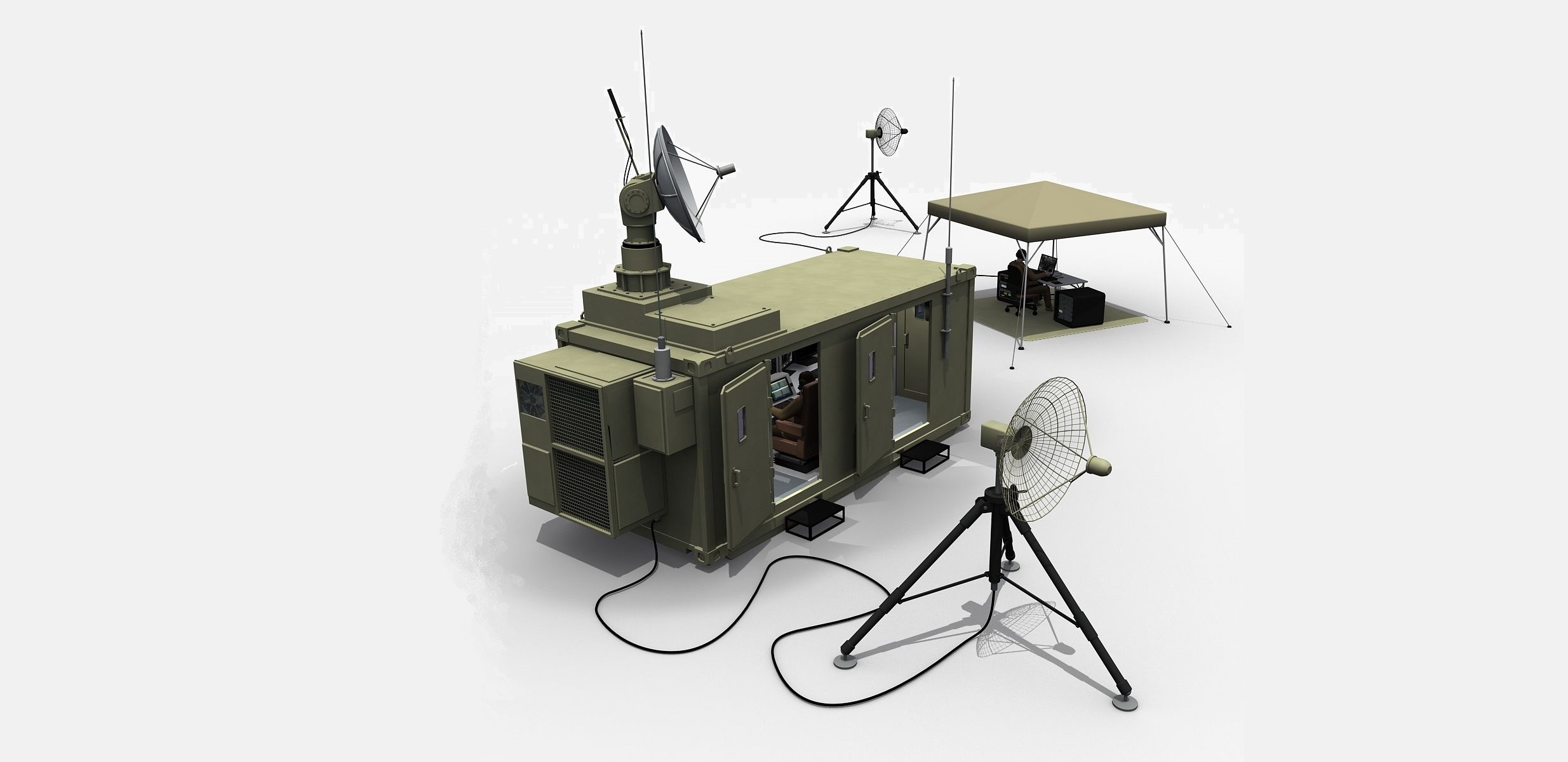}
    \caption{A 3D model for a UAV control ground station demonstrating a mobile ground control station with pilot and internal consoles and an outside control station~\cite{UAV_station_model}.} 
\end{subfigure}
\vspace{-0.5em}
\caption{The UAV Ground Control Station Concept.}
\label{fig:UAV_groundStation}
\vspace{-1em}
\end{figure}

\subsubsection{Military-Specific Constraints for UAV-Based Aerial Relays}
UAV-based aerial relays in military operations face constraints that extend well beyond civilian deployments. These constraints arise from contested environments, adversarial interference, and the need for mission assurance. The most critical factors include:
\\
\begin{enumerate}

    \item \textit{Electronic Warfare Threats:}      
    While UAV relays provide valuable EW support, they are themselves highly vulnerable to adversarial EW. Military EW targets their communication links, navigation systems, and control mechanisms, exploiting the electromagnetic spectrum to disrupt relay operations \cite{yu2025uavEW,vgi2023dronesEW}. The main
     EW threats include \cite{yu2025uavEW}:

\begin{itemize}
    \item GPS Spoofing: Adversaries transmit counterfeit GPS signals that overpower genuine satellite signals, forcing the UAV to compute false positions or times, potentially diverting it off course or into threat zones.
    
    \item Communication Jamming: Wideband, barrage, or reactive jamming reduces the SNR of C2 or TDL channels, severing operator control and disrupting inter-UAV coordination.
    
    \item Deceptive Signal Injection: Advanced EW systems can imitate valid TDL messages and inject fabricated data, creating phantom nodes, false targets, or misleading positional information that corrupts the network's operational picture.
    
    \item Carry-Off Attacks: An adversary synchronizes and gradually amplifies a counterfeit signal until the UAV locks onto it, enabling manipulation of navigation or TDL data and spreading misinformation across the relay network.
    
    \item Eavesdropping: Intercepting unencrypted or weakly protected UAV transmissions to gain intelligence or identify network vulnerabilities.
    
    \item Directed-Energy Attacks: High-power microwave (HPM) or directed-energy weapons (DEWs) aimed at disabling or damaging UAV electronics.
    
    \item Cyber Intrusions: Malware, hardware Trojans, or ground-station impersonation attempts to hijack UAV control, disrupt routing, or compromise relay integrity.
\end{itemize}

  These threats underscore the necessity for anti-jamming measures, secure communication protocols, resilient navigation systems, and cyber-hardened designs to ensure reliable military aerial-relay operations.

    \item \textit{Low Probability of Intercept/Detection (LPI/LPD):} Low probability of intercept/detection (LPI/LPD) is vital for military UAV relay operations, ensuring that communication signals remain undetectable and resilient against adversary electronic surveillance. UAV relays extending battlefield networks must maintain secure links to support mission-critical C2, reconnaissance, and coordination tasks. While LPI and LPD are often mentioned together, they differ significantly, with LPD providing a greater advantage in tactical scenarios. Although perfect LPD is unlikely, emerging technologies using the V-band offer intrinsically robust LPD, leveraging atmospheric oxygen to block transmissions at a distance \cite{armytech2025LPD}. Covert operation further requires directional beamforming, adaptive power control, and frequency hopping to minimize detectability while sustaining reliable links. Power optimization is particularly important in military scenarios, as transmitting at the minimum necessary power reduces the risk of interception while ensuring the required quality of service. By carefully managing transmission power, link reliability, and energy consumption, UAV relays can achieve stealthy, reliable, and mission-aligned communications in contested and hostile environments.
    
    \item \textit{Cyber Resilience:} Contested environments expose UAV networks to hacking, malware, signal manipulation, and denial-of-service attacks. Relays must integrate encryption, secure communication protocols, and autonomous threat detection to ensure network integrity \cite{9034071}.
    \item \textit{Integration with Military Networks:}
Integrating UAV relays with military applications differs from civilian use because of the unique operational and threat environments. In military, UAV relays are critical for connecting diverse tactical data links, such as Link-16 and Link-22, with other C2 systems, enabling a common operational picture and extending communication range in contested electromagnetic environments \cite{8493134}. Military integration requires robust cyber resilience through encryption and agile network reconfiguration, careful management of limited bandwidth to prioritize mission-critical data, and adaptive handling of dynamic network topologies caused by UAV mobility. Additionally, UAV relays must operate effectively under EW threats, employing LPI/LPD techniques and anti-jamming measures to maintain reliable connectivity in contested battlespaces.
 \item \textit{Endurance and Survivability:} 
 
 Military UAV-based aerial relays must balance mission duration, payload limitations, and exposure to enemy countermeasures to ensure reliable operation under hostile conditions. Limited endurance can constrain persistent coverage, while susceptibility to detection, interception, or kinetic threats increases operational risk. These factors require careful relay deployment planning, prioritization of critical communication links, and the use of adaptive measures such as swarm redundancy, rapid repositioning, and autonomous recovery to maintain connectivity in contested battlespaces.
 
\end{enumerate}
These points demonstrate that military UAV relays require purpose-built designs prioritizing survivability, security, and operational adaptability, rather than simply repurposing civilian communication technologies.

\subsubsection{Aerial Relay Systems in Action: Practical Examples}
   
   The ScanEagle, developed by Insitu, is utilized by military and civilian organizations alike, acting as a relay system for video and sensor data during maritime operations~\cite{scaneagle2024}. Another example, is the AeroVironment RQ-20 Puma, which provides critical situational awareness on the battlefield by relaying video and sensor data to operators. The X-47B, an experimental unmanned combat air vehicle for the U.S. Navy, has demonstrated its capability as a communication relay in naval operations. Additionally, the Bell V-247 Vigilant, designed for vertical takeoff and landing operations, can serve as a relay platform for maritime surveillance, extending communication range for naval forces. 
   
   The DA62 MPP, a multi-purpose platform, and the Boeing Insitu Integrator are also capable of relaying data and video feeds to ground control stations in various roles. Lastly, Hood Technologies' UAVs are employed in military exercises and training operations, facilitating data transmission in challenging environments.

In Fig.~\ref{fig:UAV_groundStation}, we show a typical field deployable control ground station and command center for UAVs, which provides multiple functionalities for UAVs such as airfield communication, weather station and Flight Instrumentation System.

\subsection{Aerial RIS (ARIS)}	 

\begin{figure}[t]
    \centering
    \includegraphics[width=\columnwidth]{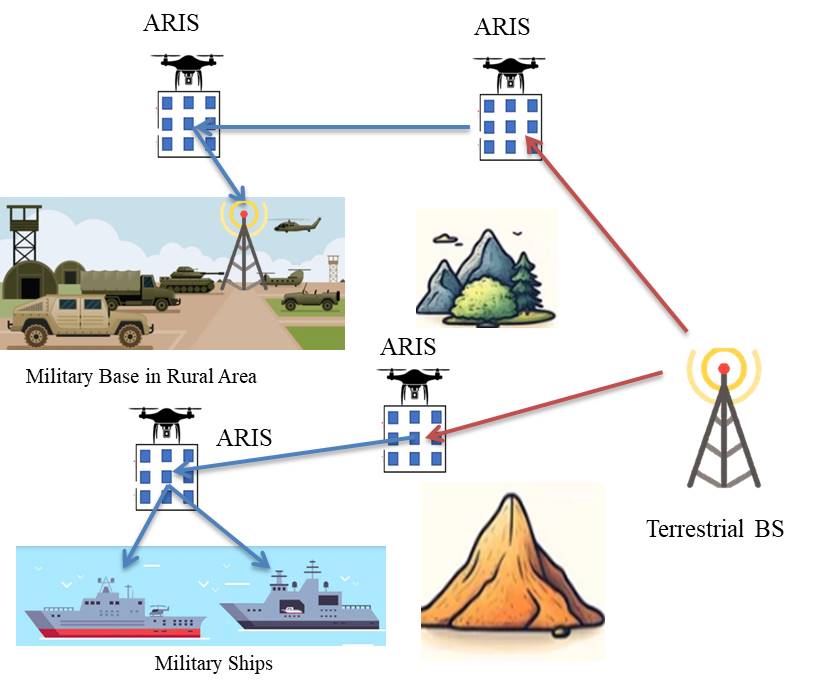} 
    \caption{UAV as an aerial RIS relay.} 
    \label{fig:UAV as an aerial RIS relay}
    \vspace{-1em}
\end{figure}

 \begin{figure*}[t]
    \centering
    \includegraphics[width=0.8\textwidth]{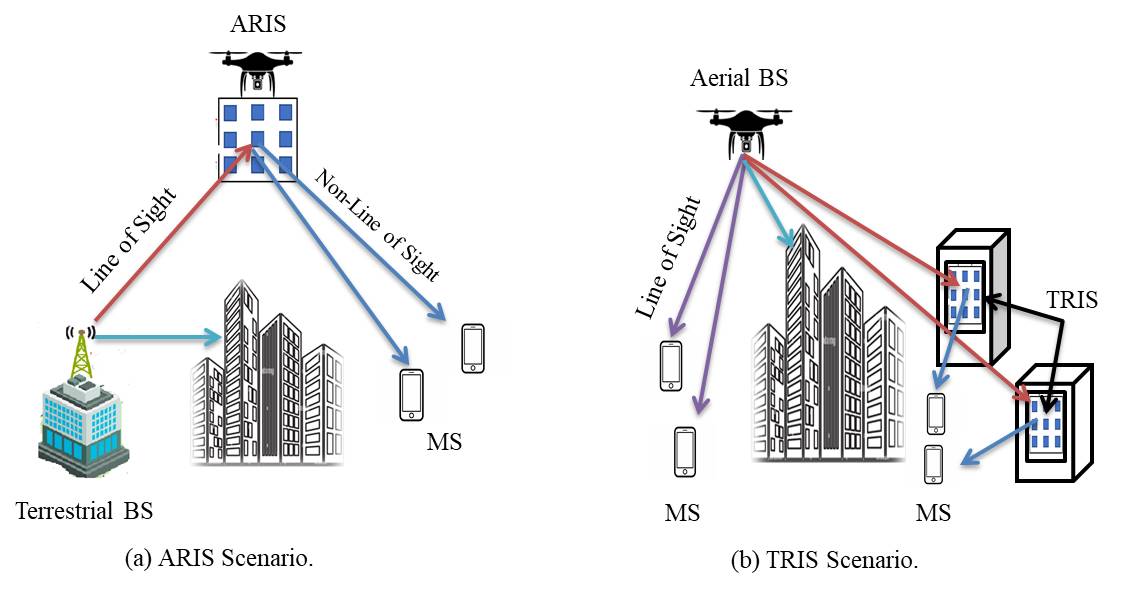} 
    \caption{RIS-assisted UAV wireless communications.}
    \label{fig: RIS types}
    \vspace{-1em}
\end{figure*}

 An ARIS system enhances communication by dynamically reflecting and directing signals. This system not only leverages the advantages of both UAVs and RIS but also mitigates their limitations. For example, low-cost UAVs face constraints in battery life and bandwidth, making them unsuitable as small-cell flying base stations. However, integrating a RIS panel on the UAV can greatly lower energy consumption, addressing these challenges.
 
 Rather than frequently repositioning the UAV to adjust to changes in network topology caused by UEs' mobility, we can modify the phase shifts of the RIS to establish new LoS links, eliminating the need for continuous UAV movement~\cite{9810504}. Fig.~\ref{fig:UAV as an aerial RIS relay} depicts UAVs equipped with ARIS functioning as relays to extend coverage. In this figure, the deployment of multiple ARIS is utilized to extend coverage over long distances, thereby improving communication with a military base in rural areas and with military ships at sea.

\subsubsection{Definition and Operation}
RIS adjusts the signal environment without active signal processing by using passive elements that manipulate the phase, amplitude, and polarization of incoming electromagnetic waves. These elements, often made of meta-materials, can be dynamically tuned to create specific interference patterns, directing and focusing the reflected signals toward desired directions through coordinated phase shifts, a process known as beamforming. This allows RIS to enhance signal quality, extend coverage, and reduce interference without the need for energy-intensive amplification or signal regeneration~\cite{8936989}.
  
Controlled by an external system, RIS operates with low power consumption and complexity, making it an efficient and scalable solution for improving signals~\cite{10261452, 9703337, 10208239}. When RIS technology is mounted on aerial platforms like drones or UAVs, it is called ARIS. In contrast, when RIS is placed on terrestrial platforms such as towers or buildings, it is known as terrestrial RIS (TRIS)~\cite{8959174}. Fig.~\ref{fig: RIS types} shows both ARIS and TRIS. ARIS offers greater flexibility and coverage by moving in 3D space to optimize signal reflection, unlike stationary TRIS~\cite{9434412}. It can adjust its position in real-time to target specific areas or devices, and its elements can be individually controlled to steer signals using beamforming techniques~\cite{9141340}. This dynamic positioning allows ARIS to improve connectivity and adapt to environmental changes for reliable communication~\cite{10261452}.

In modern relay networks, characterized by intense wave propagation, strong interference, and heavy RF jamming due to the integration of intelligent weapon systems with C2 networks, maintaining coverage is essential for uninterrupted line-of-sight and long-distance communication~\cite{globalhawk2024, predator2024}. RIS offers an effective solution to these challenges by enhancing spectral and energy efficiency, enhancing signal strength and minimizing interference~\cite{scaneagle2024, 9508885}. This enhances coverage and ensures reliable long-distance communication, even in complex environments where weapon systems are integrated.

RIS uses passive elements that can adjust the amplitude and phase of signals to improve link quality and coverage~\cite{9374975, 9159923, 8968350}. Compared to active relays, RIS is more energy-efficient, operates in full-duplex mode without self-interference, and allows for a controllable channel propagation environment~\cite{8910627, 8466374}. In \cite{yao2023}, it was reported that RIS-assisted UAVs can enhance energy efficiency by up to 145\% compared to AF relays.

Recent studies have focused on integrating UAVs into next-generation communication systems, particularly for overcoming communication challenges with RIS applications. For example, \cite{8959174} proposes using RIS to mitigate line-of-sight loss due to environmental factors in UAV trajectories, while \cite{9120632} demonstrates signal enhancement with RIS placed on building walls.

Comparative analyses and optimizations of RIS-UAV systems are explored in~\cite{adam2024secure, 9351782, 9124704}, addressing aspects like system performance, coverage, and outage probability. Additionally, \cite{zhao2021outage} examines network performance with mobile users, and \cite{9293155} focuses on rate maximization in OFDM-based UAV systems. Furthermore, \cite{zhou2021aerial} discusses Doppler effect and power optimization in RIS-supported massive MIMO systems, while \cite{9462487} looks into path and phase shift optimization.

The studies in~\cite{zhao2022simultaneously} and \cite{9416239} present methods for enhancing secure communication and signal direction using advanced algorithms and RIS structures. In \cite{9699402}, it was found that low-complexity RIS-supported UAVs significantly improve performance in high-mobility communications compared to non-RIS designs. In \cite{9599478}, the effects of interference in airborne RIS-assisted inter-vehicular communication scenarios were analyzed. Additionally, \cite{9822386} demonstrated that UAVs with RIS can significantly enhance communication performance in challenging fading channels. The authors in \cite{9989438} proposed using a UAV-mounted RIS for data collection, evaluated its coverage probability, and introduced a new medium access control protocol. 

The study in~\cite{9777886} explored a UAV swarm-enabled ARIS (SARIS) system for maximizing ground user sum-rate, proposed low-complexity beamforming and optimal SARIS placement, and validated its effectiveness at larger distances through simulations.
The results of this study show that multi-user beamforming achieves a 1-30\% gain, and placement optimization enhances SARIS by 20-500\% compared to the scenario without placement optimization. Additionally, optimal placement depends on user distance, and NLoS estimation can be skipped with minimal performance loss.

The authors of \cite{10289638} integrated RIS and UAV into a wireless-powered network, enabling IoT devices to communicate using energy harvested from an energy station. The research work in \cite{10494543} addressed secure communication in multiuser networks using UAV-mounted RIS, optimizing beamforming, phase shift, and UAV trajectory to maximize secrecy rates under channel uncertainty. The study in \cite{10668867} introduced an adaptive UAV swarm and dynamic device clustering technique with RIS to optimize interference management in smart cities using Reinforcement Learning. The authors in \cite{9277627} integrated UAV-RIS with NOMA to minimize power consumption, achieving a 23.3\% reduction in UAV energy usage and 11.7\% compared to OMA systems. The work in \cite{10449453} analyzes UAV-mounted RIS systems by deriving outage probability, ergodic capacity, and energy efficiency expressions, and demonstrates their advantages over decode-and-forward relaying in composite fading environments. Finaly, the authors of \cite{10854532} propose a block coordinate descent (BCD)-based optimization scheme for multi-RIS-assisted UAV networks, deriving closed-form SNR expressions and jointly optimizing UE-RIS-UAV association, placement, and phase alignment. Their approach demonstrates 12–45\% connectivity gains over benchmarks, highlighting its efficacy for massive connectivity in B5G networks.

While these studies collectively highlight the versatility of RIS and UAV integration in improving coverage, energy efficiency, and secure communication, most evaluations remain confined to controlled or idealized conditions. Critical factors such as mobility under adversarial jamming, real-time reconfiguration in contested airspaces, and scalability for large UAV swarms are often overlooked. For military applications, these limitations are non-trivial: centralized control and reliance on perfect channel knowledge may be infeasible in EW environments where GPS signals are spoofed, communication links are degraded, or UAV losses occur. Thus, despite the aforementioned advantages, further research is required to investigate decentralized coordination, resilience against jamming, and rapid reconfigurability in order to address the unique demands of military missions.

\subsubsection{UAV Size and RIS Integration}

The relationship between UAV size and the integration of RIS is crucial for optimizing communication system performance. Larger UAVs offer significant advantages in terms of payload capacity, allowing them to support larger and heavier RIS units. This increased capacity enables the integration of more extensive RIS panels, which can enhance signal coverage and quality more effectively.

Larger UAVs also tend to have better stability, which helps maintain accurate positioning of the RIS, further improving its performance. Additionally, these UAVs can handle the added weight of the RIS without substantially impacting their flight dynamics, thereby preserving operational range and efficiency. Conversely, smaller UAVs face limitations when integrating RIS due to their restricted payload capacity. The size and weight constraints of smaller UAVs mean that the RIS must be more compact, which can limit its effectiveness in signal enhancement and coverage. 

The additional weight of even a small RIS can affect the UAV's stability and reduce its flight time, impacting overall performance. The design and integration of RIS on smaller UAVs must be carefully managed to avoid negatively affecting the UAV's flight dynamics and energy consumption. Furthermore, the design flexibility for integrating RIS is greater with larger UAVs. They offer sufficient space for experimenting with various RIS configurations and placements to optimize performance.

In contrast, integrating RIS into smaller UAVs requires precise design to avoid adverse impacts on the UAV's stability and flight characteristics. Installation and maintenance are also more straightforward on larger UAVs due to the available space, whereas smaller UAVs may present challenges in these areas, demanding more meticulous integration and maintenance efforts. 
    
In summary, the size of the UAV significantly influences the effectiveness of RIS integration. Larger UAVs provide enhanced capability for deploying RIS, leading to better performance and coverage. On the other hand, smaller UAVs are constrained by payload limitations and operational efficiency, which can affect the RIS’s effectiveness. Balancing these factors is crucial for optimizing UAV-based communication systems.

\subsubsection{ARIS Applications}

ARIS can enhance communication system performance in both civilian and military applications. The key distinction between the two lies in their operational environments and design priorities. Civilian systems can benefit from ARIS through improved coverage, better spectral efficiency, and enhanced connectivity in hard-to-reach areas. In contrast, military communications demand secure, reliable, and adaptive performance in dynamic, hostile, and infrastructure-limited settings. Therefore, in military scenarios, UAV-mounted RIS can play a crucial role by acting as an agile, mission-driven platform capable of real-time repositioning. It can improve signal strength and enables secure communication through passive beam steering, reducing the risk of jamming and interception. 

Recent studies show that RIS significantly enhances the secrecy performance of UAV-assisted wireless systems compared to those without RIS. For example, the study in \cite{9367542} demonstrate a 38\% improvement in uplink secrecy energy efficiency using joint optimization of UAV trajectory, RIS phase shifts, user association, and power. Similarly, \cite{9209992} shows that optimizing UAV trajectory, power control, and RIS configuration significantly improves secrecy rates, especially with carefully chosen phase shifter counts and increased reflecting elements. Furthermore,the authors of \cite{10070838} design a UAV-mounted RIS scheme to enhance secure transmission by jointly optimizing UAV position, beamforming, and RIS phase shifts, demonstrating significant security improvements through simulation results. The authors of \cite{10146504} propose a UAV-mounted multi-functional RIS that reflects, amplifies, and emits jamming signals to enhance security. Simulation results in \cite{10146504} demonstrate that the proposed approach significantly improves secrecy performance compared to existing approaches. The authors of \cite{9538830} analyze the secrecy performance of an ARIS relay under various system parameters and demonstrate its effectiveness. In \cite{10214219},  the authors propose a power-efficient secure transmission scheme and confirm the performance advantages of active RIS. In \cite{10283889, 10345491}, the authors apply deep reinforcement learning to jointly optimize UAV trajectory, beamforming, and RIS settings to maximize secrecy under QoS and reliability constraints. Simulation results presented in \cite{10283889, 10345491} indicate that the proposed methods offer a notable improvement in the secrecy rate over the existing methods.

These studies collectively highlight the importance of combining UAV and RIS to improve security in UAV-assisted systems. The ability of UAV-mounted RIS to enhance communication security, improve signal strength, and reduce interception risks makes it essential for secure and reliable operations in military applications.

\subsection{Technological Advancements in Aerial Relays}
    
Recent technological advancements in UAVs have significantly improved their performance, allowing for enhanced ARs capabilities that optimize connectivity and operational efficiency across various applications, including military and civilian. Here are some of these advancements:

\begin{enumerate}[label=\alph*)]
    \item \textit{Integration of UAVs with 5G and Beyond:} 
Integrating UAVs as flying base stations or relays represents a significant advancement in 5G and beyond, and this topic has recently attracted considerable interest ~\cite{10379625, 8579209, khan2021role, 9768113, 10208239}. The survey in ~\cite{10379625} discusses the applications of UAVs as base stations and their potential to influence the future of communication technologies, including 5G and beyond. In ~\cite{8579209}, the authors discuss how UAVs enable wireless broadcasting and high-rate transmissions, addressing the challenge of providing ubiquitous connectivity for diverse devices in 5G and beyond. In ~\cite{khan2021role}, the authors explain that densified 5G networks provide higher throughput and lower latency, with UAVs identified as a practical solution for their on-demand deployment. In ~\cite{9768113}, the tutorial explores the evolving role of UAVs in cellular communications, covering the transition from 5G to 6G. The paper presents several case studies and original results, offering insights into how UAVs will be integral to next-generation networks by improving mobility, spectrum efficiency, and resource management. Furthermore, in ~\cite{10208239}, the authors discuss how the energy efficiency and capacity of the 5G network can be improved by employing UAVs and RIS.

\item \textit{AI Integration and Autonomous UAV Operations:}
Advances in artificial intelligence (AI) and machine learning (ML) have significantly enhanced UAV autonomy, enabling real-time decision-making, complex navigation, and minimal human intervention~\cite{9732774,Bithas2019}. Studies show that ML techniques improve wireless networks by optimizing coverage, enhancing data transmission, and enabling UAVs to act as adaptive aerial relays~\cite{9220178,10113154,ijgi9010014,LIU2020253,SRIVASTAVA2021102152,9439930,Osco2021,Iftikhar2023}.

Practical implementations, such as the Skydio X2 UAV, use AI algorithms for obstacle avoidance and mission planning, improving safety and efficiency~\cite{skydio2024}. In military contexts, AI-driven UAVs enable adaptive ISR missions, resilient relay-based communications, and dynamic mission planning in contested environments.

Beyond single platforms, multi-UAV networks expand mission capacity but demand intelligent coordination. Reinforcement learning (RL) offers solutions by optimizing trajectories, resource allocation, and relay placement in real time~\cite{9768113,10283826,10494323,10531095,9778241}. Collectively, these advances highlight UAVs as intelligent, autonomous, and adaptive relays, critical for both civilian and defense operations.
     
 \item \textit{UAV Swarms:} Multi-UAV networks, known as UAV swarms, have recently gained significant attention for their ability to perform complex missions more efficiently ~\cite{10283826, 10430396}. They provide enhanced intelligence, better coordination, increased flexibility, higher survivability, and adaptability. These systems require the integration of various components, including path planning, localization, and task coordination ~\cite{10430396}. UAV swarms can enable multiple ARs to operate in coordination, sharing information and resources in real-time. This capability can enhance the effectiveness of operations such as search and rescue, surveillance, and military applications by improving coverage and responsiveness~\cite{10430396}.
\item \textit{Integration with IoT:} 
IoT has been a key technology for a decade, with significant research showing its crucial role in UAVs for enhancing data transmission between IoT devices and UAVs. This enables ARs to connect seamlessly with IoT devices, e.g., sensors installed in agricultural fields to monitor soil conditions, creating a networked environment that enhances data collection and analysis~\cite{9703337, abbas2023,8995526, 9386233}. In~\cite{abbas2023}, the authors explore how smart cities use IoT and UAVs to enhance residents' quality of life. UAVs can also adjust their positions based on IoT patterns, enhancing connectivity and energy efficiency without requiring extensive ground infrastructure, making them a key solution for IoT communication challenges ~\cite{10379625}.
The work in  ~\cite{8995526} discusses the role of UAVs in IoT and introduces a model to optimize their positioning. In ~\cite{9386233}, the authors use a  UAV-mounted RIS to enhance data transmission and improve spectral efficiency in IoT networks, with a BS-UAV and UAV-user link, and no direct link between the BS and user. 

\item \textit{Enhanced Stealth and Survivability Technologies:}
Stealth and survivability technologies are critical for UAVs operating in contested environments. Advances in materials and design have led to UAVs with reduced radar cross-section (RCS) and infrared signatures, making them less detectable by enemy sensors.
RCS or "echo area" of a UAV is a measure of how detectable it is by radar. It quantifies the amount of electromagnetic energy that is reflected back to the radar receiver when the radar signal hits the UAV, and it is defined 
as~\cite{balanis2015antenna}
\begin{align}\label{eq:RCS}
\sigma = 
\lim_{R \rightarrow \infty}
\left[
\frac{4 \pi R^2 W_s}{W_i}
\right]
\quad (\text{measured in}\ m^2)
\end{align}
where $R$ is the distance between the UAV and the radar measured in~$m$, and $W_i$ ($W_s$) is the incident (scattered) power density which is measured in~$W/m^2$.

The incident and scattered power density can be directly related to the incident and scattered electric fields ($E_i$ and $E_s$) and magnetic fields ($H_i$ and $H_s$) as
\begin{align}
\frac{W_s}{W_i}
=
\frac{|E_s|^2}{|E_i|^2}
=
\frac{|H_s|^2}{|H_i|^2}
\end{align}

Using the definition of the RCS in~\eqref{eq:RCS}, we can calculate the amount of power captured by a radar for signals that hit the UAV and re-radiated isotropically back. This amount of power captured can be represented as

\begin{equation}
P_r = \frac{P_t G_t G_r \lambda^2 \sigma}{(4 \pi)^3 R^4},
\end{equation} \\
where $P_t$ is the transmitted power, $G_t$ is the transmitter antenna gain, $G_r$ is the receiver antenna gain, $\lambda$ is the signal wavelength, $\sigma$ is the radar cross-section, and $R$ is the distance between the UAV and the radar.

As a practical example, the X-47B, developed by Northrop Grumman for the U.S. Navy, is a tailless combat UAV (UCAV), designed for stealth~\cite{NorthropGrumman2024}. The Northrop Grumman X-47B features stealth technology that minimizes its radar profile, enhancing its ability to perform reconnaissance and strike missions in high-risk areas. The Lockheed Martin RQ-170 is another example~\cite{AirForce2024}. It features stealth technology to minimize its RCS, enhancing its survivability in hostile airspaces. The flying wing design further reduces radar visibility. 

Additionally, innovations in UAV design and materials are enhancing performance, durability, and mission versatility. New lightweight composites and aerodynamic designs improve UAV efficiency and endurance while reducing overall weight. The Lockheed Martin RQ-170 Sentinel, for example, utilizes advanced materials to achieve a balance between stealth and operational range. These advancements collectively drive the evolution of military UAVs, enhancing their effectiveness, adaptability, and resilience in modern warfare.

The precise RCS for military UAVs is often classified or not publicly disclosed due to security considerations. However, the \textbf{threshold for stealth classification} is typically $\le 0.1~{\rm m}^2$. UAVs such as RQ-170 Sentinel, X-47B, and BAE Systems Taranis, fall into this category and are considered stealth UAVs. These UAVs achieve low RCS through advanced design techniques and the use of radar-absorbing materials (RAM) which have the ability to absorb electromagnetic waves. For comparison, the RCS of common objects varies significantly, where a human has an approximate RCS of $1~{\rm m}^2$, a bird has an RCS of $0.01~{\rm m}^2$, and an insect has an RCS of $10^{-5}~{\rm m}^2$. In Fig.~\ref{fig:rcs_plots}, we provide the RCS of typical objects.
    
\end{enumerate}

\begin{figure}[t]
    \centering
    \includegraphics[width=1.15\linewidth]{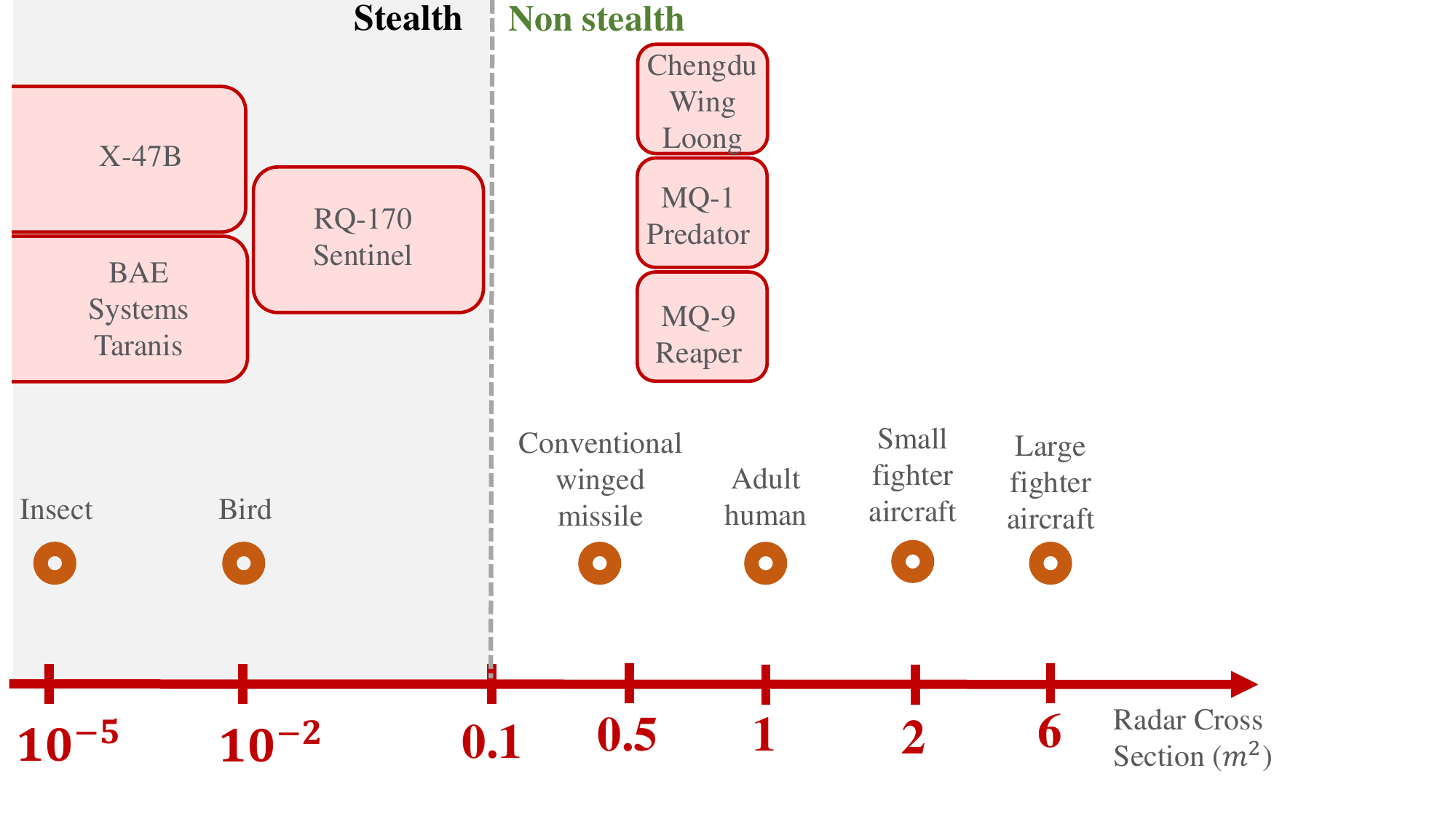}
    \vspace{-2em}
    \caption{RCS of typical objects.} 
    \label{fig:rcs_plots}
    \vspace{-1em}
\end{figure}

\begin{table*}[htbp]
\caption{Comparative Analysis of Relay Types}
\centering
\scriptsize
\fontsize{9}{9}\selectfont
\begin{tabular}{@{}p{3cm}p{2.8cm}p{3.4cm}p{3cm}p{2.8cm}@{}}
\toprule
\textbf{Parameter} & \textbf{AAR} & \textbf{ATR} & \textbf{ARIS} & \textbf{TRIS} \\ 
\midrule

\multicolumn{5}{l}{\textbf{1) Mobility}} \\
Mobility & High & None (fixed) & High & None (fixed) \\

\midrule
\multicolumn{5}{l}{\textbf{2) Jamming Resilience}} \\
Resilience & Moderate & High & High & High \\
\midrule
\multicolumn{5}{l}{\textbf{3) Deployment}} \\
Deployment Speed & Fast  & Slow  & Fast  & Slow \\

\midrule
\multicolumn{5}{l}{\textbf{4) Stealth \& Security}} \\
Detectability & Low & Moderate & Low & Low \\
Survivability & Moderate & High  & Moderate  & High \\
\midrule
\multicolumn{5}{l}{\textbf{5) Coverage \& Performance}} \\
Coverage Area & Wide & Widest and stable & Medium & Small \\
Range & Long  & Long & Moderate & Short \\
Signal Quality & High  & High  & Moderate & Moderate to Low \\
Latency & Low & Low & Very Low (reflection) & Very Low (reflection) \\

\midrule
\multicolumn{5}{l}{\textbf{6) Autonomy \& Sustainability}} \\
Autonomy Level & Medium & High & Medium & High \\

Endurance & Limited & Long & Limited & Long \\
Power Use & High & Low  & Moderate  & Very Low \\
Envelope Resilience & Low  & High  & Low  & High  \\
Maintenance & High  & Moderate  & High  & Low  \\
Payload / Drag & Limited payload, drag-sensitive & High payload, no drag issue & Limited payload, drag-sensitive & High payload, no drag issue \\

\bottomrule
\end{tabular}
\label{table:relay_RIS_comparison}
\end{table*}

\subsection{Lessons Learned} 
The analysis of aerial relays (ARs) yields several concrete insights for military communication network design. First, AAR provide strong coverage expansion and throughput gains—up to 47\% in dynamic and 40\% in static missions—making them ideal for expeditionary and long-range tactical operations. However, these advantages come with increased energy consumption and maintenance demands, elevating logistical risk and mission cost. Second, ARIS offer a complementary low-power solution that enhances energy efficiency, stealth, and electromagnetic concealment, making them particularly suited for covert or infrastructure-denied environments, though at the expense of limited processing capability and adaptability under mobility. Third, endurance and survivability remain central constraints: AARs face operational limits due to power dependence, while ARIS performance degrades under severe weather and alignment errors. Finally, no single relay architecture dominates across all metrics. Instead, hybrid relay configurations—combining AARs’ active gain with ARIS’ energy-efficient reflection—represent a promising path toward resilient, cost-effective, and adaptable communication networks capable of sustaining secure command and control in contested airspaces.

\section{Comparative Analysis}

This section begins with a qualitative comparison in Subsections V-A to V-C, examining relay types across six mission-critical dimensions—mobility, jamming resilience, deployment speed, stealth and security, coverage and performance, and autonomy and sustainability—as summarized in Table~\ref{table:relay_RIS_comparison}. 
The selection of these six features reflects established doctrine-level priorities and translates them into measurable operational requirements. Mobility and deployment speed capture readiness and tactical agility, which are central to capability-development processes as outlined in the joint capabilities integration and development system (JCIDS) \cite{JCIDS}. Jamming resilience and stealth/security address the need to operate in contested electromagnetic environments while minimizing electromagnetic signatures that increase detection risk, as emphasized in DoD Instruction 4650.01 on electromagnetic spectrum operations \cite{DoDI4650}. Coverage/performance represents core requirements of allied joint publication (AJP)-6 on communications and information systems (CIS), including range, reliability, and interoperability needed to sustain C2 and user communications \cite{AJP6}. Finally, autonomy/sustainability encompasses endurance, logistical footprint, and life-cycle supportability, as highlighted in MIL-HDBK-502 on Product Support Analysis \cite{MILHDBK502}.

Building upon this qualitative foundation, Subsection V-D introduces our core contribution: a novel decision-support framework centered on the Mission-Critical Relay Effectiveness Score (MCRES). This multi-dimensional metric quantifies relay performance by integrating the six doctrine-derived parameters with mission-specific weights. The metric is used within a structured selection framework, formalized in Algorithm 1, which provides a step-by-step procedure for inputting scenario data, computing scores, and determining the optimal relay type.
\\
\subsection{Comparing AAR and ATR}  
The comparison between AAR and ATR illustrates a fundamental trade-off between mobility and stability. As presented in Table~\ref{table:relay_RIS_comparison}, AAR enables rapid deployment and flexible repositioning, making it particularly valuable in highly dynamic contexts such as disaster response, tactical battlefield communications, and temporary hotspot coverage. Its altitude advantage mitigates line-of-sight obstructions, thereby extending coverage and facilitating faster adaptation to evolving operational requirements. However, these benefits are counterbalanced by significant limitations, including elevated operational costs, restricted endurance due to propulsion and energy demands, limited payload capacity arising from drag sensitivity, and reduced survivability under severe weather or adversarial conditions.
\\
By contrast, ATR ensures long-term reliability and cost-effectiveness through fixed terrestrial infrastructure. Continuous access to stable power supply, higher payload capacity, reduced maintenance demands, and greater environmental resilience render it more suitable for sustained missions. Although ATR lacks mobility and may be more susceptible to jamming attacks given its stationary nature, its superior survivability and operational persistence make it well aligned with enduring applications such as border surveillance, permanent military installations, and civilian infrastructure support.
\\
In summary, AAR is most appropriate for short-duration, high-mobility missions requiring rapid adaptability, whereas ATR is better suited to long-term deployments where efficiency, survivability, and reliability are paramount.
\\

\subsection{Comparing ARIS and TRIS}  
ARIS integrates UAV-mounted RIS, combining aerial mobility with passive reflection to enhance adaptability across dynamic missions. Its high mobility and rapid deployment speed  make it particularly effective in urban disaster relief or military operations where shadowed areas and rapidly evolving conditions demand quick reconfiguration. ARIS also provides strong coverage flexibility, enabling temporary extension of communication links into remote or obstructed environments. However, these advantages are offset by limited autonomy due to UAV battery life, reduced survivability under adverse weather conditions, and increased payload/drag considerations that restrict endurance.  

Conversely, TRIS offers stable and energy-efficient  performance, with high survivability and low operational overhead. Its fixed nature ensures strong jamming resilience and reliable stealth and security, making it well suited for long-term, infrastructure-supported missions such as base protection or smart city monitoring. Nonetheless, TRIS lacks mobility and adaptability, which can create coverage gaps in rapidly changing or geographically dispersed scenarios.  

The last two columns of Table~\ref{table:relay_RIS_comparison} summarize these characteristics, highlighting ARIS as the preferred solution for dynamic and hard-to-reach operations, while TRIS is more suitable for stable, enduring deployments requiring efficiency and resilience.

\begin{table*}[!t]
\centering
\caption{Suggested Weights \(w_i\) per Scenario  }
\label{tab:scenario_weights_justified}
\begin{tabular}{p{2.7cm} c c c c c c p{8.3cm}}
\toprule
\textbf{Scenario} & \(\boldsymbol{w_M}\) & \(\boldsymbol{w_J}\) & \(\boldsymbol{w_D}\) & \(\boldsymbol{w_S}\) & \(\boldsymbol{w_C}\) & \(\boldsymbol{w_A}\) & \textbf{ Justification} \\
\midrule
Dynamic Battlefield & 0.30 & 0.25 & 0.25 & 0.20 & 0.00 & 0.00 & 
\textbf{JCIDS} emphasizes tactical agility through mobility and deployment speed \cite{JCIDS}. \textbf{DoDI 4650.01} mandates resilience in contested EMS. Stealth critical for survivability in contact. Coverage treated as baseline enabler \cite{DoDI4650}. \\
\hline
Search \& Rescue & 0.40 & 0.00 & 0.40 & 0.00 & 0.20 & 0.00 & 
\textbf{NATO SAR procedures} prioritize rapid area coverage and deployment speed for casualty survival. Connectivity secondary but essential for coordination. Jamming unlikely in humanitarian scenarios \cite{ATP10D}. \\
\hline
Electronic Warfare & 0.10 & 0.40 & 0.00 & 0.40 & 0.10 & 0.00 & 
\textbf{DoDI 4650.01} prioritizes spectrum operations in contested environments. Primary survivability metrics are jamming resilience and stealth/security \cite{DoDI4650}. Mobility secondary for emitter avoidance. \\
\hline
Covert Operations & 0.00 & 0.25 & 0.00 & 0.50 & 0.00 & 0.25 & 
Military stealth doctrine emphasizes signature reduction as paramount. Resilience prevents detection via emissions. Autonomy ensures mission completion without resupply. Mobility could compromise position. \\
\hline
Disaster Response & 0.30 & 0.00 & 0.40 & 0.00 & 0.20 & 0.10 & 
\textbf{FEMA response protocols} emphasize rapid deployment and mobility in damaged infrastructure. Coverage essential for coordination. Limited endurance required for sustained operations \cite{FEMA_NRF}. \\
\hline
Long-Endurance Surveillance & 0.00 & 0.10 & 0.00 & 0.20 & 0.20 & 0.50 & 
Prioritizes persistent and covert monitoring, requiring high sustainability, stealth for operational security, and reliable coverage, while mobility remains a secondary concern. \\
\hline
Urban Base Comm. (new) & 0.00 & 0.20 & 0.00 & 0.20 & 0.20 & 0.40 & 
High autonomy, jamming resilience, and stealth are required for persistent, secure operation. \\
\hline
Fixed Base Comm. (new) & 0.00 & 0.20 & 0.00 & 0.20 & 0.20 & 0.40 & 
High autonomy, jamming resilience, and stealth are required for fixed, persistent, and secure operation. \\
\bottomrule
\end{tabular}
\end{table*}

\subsection{Comparing AAR and ARIS}  
Although both AAR and ARIS are aerial relay platforms, their operational profiles differ significantly across the mission-critical parameters in Table~\ref{table:relay_RIS_comparison}. AAR, as an active relay, amplifies and forwards signals, providing superior coverage  and extended range for wide-area battlefield operations. Its robustness against jamming makes it effective in contested environments. However, these advantages come with higher energy consumption, increased payload/drag requirements, greater maintenance complexity, and reduced autonomy, all of which increase operational costs.  

In contrast, ARIS leverages passive reflection to achieve low-power operation and rapid deployment speed, with enhanced mobility enabling coverage in shadowed or hard-to-reach areas. Nonetheless, ARIS offers limited range and weaker survivability under harsh weather conditions. Its reduced coverage and moderate jamming resilience make it less suited for large-scale or highly contested missions. These characteristics position ARIS as an energy-efficient solution for localized or temporary coverage, such as IoT support, tactical surveillance, or short-duration military missions.  

In summary, AAR should be prioritized when mission success hinges on maximizing range, coverage, and resilience, whereas ARIS provides a lightweight, energy-efficient, and rapidly deployable alternative for localized or temporary missions. While the comparative analysis in Table~\ref{table:relay_RIS_comparison} provides a structured qualitative comparison of relay types, it does not fully support mission-oriented decision-making. To address this, each mission-critical parameter will be numerically scored and combined with scenario-specific weights to compute the MCRES, as detailed in the next subsection.

\begin{table*}[!t]
\centering
\caption{Relay Capabilities Across Key parameters with Justifications and Supporting Evidence}
\label{tab:relay_capabilities_justified}
\resizebox{\textwidth}{!}{%
\begin{tabular}{p{2.2cm} c p{6.0cm} c p{6.0cm}}
\toprule
\textbf{Parameter} & \textbf{AR Score} & \textbf{AR Justification \& Evidence} & \textbf{TR Score} & \textbf{TR Justification \& Evidence} \\
\midrule
\textbf{Mobility (M)} & 1.00 & Inherent 3D mobility for dynamic repositioning and threat avoidance; core UAV advantage enabling agile C2 \cite{JCIDS}. & 0.00 & Fixed or ground-vehicle-mounted; static by definition, a key limitation of terrestrial systems \cite{AJP6}. \\
\hline
\textbf{Jamming Resilience (J)} & 0.75 & Limited onboard power restricts active countermeasures (Sec. IV.A.4). & 1.00 & Grid power and infrastructure support enable robust anti-jamming capabilities \cite{DoDI4650}. \\
\hline
\textbf{Deployment Speed (D)} & 1.00 & Can be launched within minutes for immediate coverage; embodies tactical agility (Sec. IV.A; \cite{JCIDS}). & 0.00 & Deployment requires hours to days for transport and installation, resource-intensive (Sec. III). \\
\hline
\textbf{Stealth \& Security (S)} & 0.75 & Mobility supports LPI/LPD tactics, but acoustic/visual/radar signatures remain high (Sec. IV.A.5, IV.C). & 0.75 & Can be concealed/hardened with no inherent signatures, but becomes vulnerable once located (Sec. III). \\
\hline
\textbf{Coverage/Performance (C)} & 1.00 & Altitude ensures long-range LoS, overcoming terrain limits; studies report up to 47\% coverage improvement \cite{8758183}. & 0.80 & Provides stable, high-quality coverage within a fixed footprint; reliable localized service (Sec. III). \\
\hline
\textbf{Autonomy \& Sustainability (A)} & 0.60 & Endurance limited to hours; requires frequent recovery and maintenance, increasing logistical burden (Sec. IV.A.5; \cite{MILHDBK502}). & 0.90 & Continuous power supply enables long-term unattended operation with lower life-cycle costs (Sec. III). \\
\bottomrule
\end{tabular}
}
\end{table*}

\subsection{Mission-Critical Relay Effectiveness Score (MCRES)}

This subsection presents the MCRES, a quantitative tool for evaluating and selecting the most appropriate relay type for specific military missions. It integrates six key parameters—mobility ($M$), jamming resilience ($J$), deployment speed ($D$), stealth/security ($S$), coverage/performance ($C$), and autonomy/sustainability ($A$)—with mission-tailored weights to compute a single composite score. By employing a linear weighted aggregation, MCRES provides a clear, consistent, and scalable framework for decision-making. This methodology is consistent with multi-criteria decision-making (MCDM) practices commonly used in defense acquisition and system evaluation \cite{Triantaphyllou2000}. The MCRES is calculated as:

\begin{equation}
\text{MCRES} = w_M M + w_J J + w_D D + w_S S + w_C C + w_A A,
\end{equation} 
where $w_i$ $(i = M, J, D, S, C, A)$ denote the weights reflecting the relative importance of each parameter for the mission, subject to the constraint $\sum_{i ={M, J, D, S, C, A}} w_i = 1$. The weights are mission-dependent. \\
 
Table~\ref{tab:scenario_weights_justified} presents example weights for mission-critical features across diverse military scenarios. These weights ensure that MCRES reflects doctrinally grounded trade-offs rather than treating all dimensions equally. Defense acquisition frameworks such as JCIDS \cite{JCIDS} emphasize tailoring capability priorities—mobility, survivability, or sustainability—according to mission profile. For example, in EW, directives on spectrum security \cite{DoDI4650} justify higher weights for jamming resilience and stealth, while humanitarian and disaster-relief missions prioritize mobility and rapid deployment. Similarly, long-endurance surveillance requires greater emphasis on autonomy and sustainability, consistent with life-cycle and logistics guidance \cite{MILHDBK502}. Overall, the weight selection aligns with established doctrine and acquisition standards while adapting to operational context.

\begin{table}[t]
\centering
\caption{Computed MCRES for each relay type across military scenarios.}
\label{tab:mcres_computed}
\fontsize{9}{10}\selectfont
\begin{tabular}{p{4cm} p{0.6cm} p{0.6cm} p{1.8cm}}
\toprule
\textbf{Scenario} & \textbf{AR} & \textbf{TR} & \textbf{Recommended} \\
\midrule
Dynamic Battlefield          & \textbf{0.888} & 0.400 & AR \\
Search \& Rescue             & \textbf{1.000} & 0.160 & AR \\
Electronic Warfare           & \textbf{0.800} & 0.780 & AR (slightly) \\
Covert Operations            & 0.713 & \textbf{0.850} & TR \\
Disaster Response            & \textbf{0.960} & 0.250 & AR \\
Long-Endurance Surveillance  & 0.725 & \textbf{0.860} & TR \\
Urban Base Comm.             & 0.740 & \textbf{0.870} & TR \\
Fixed Base Comm.             & 0.740 & \textbf{0.870} & TR \\
\bottomrule
\end{tabular}
\end{table}

Table~\ref{tab:relay_capabilities_justified} compares the capabilities of ARs and TRs across six mission-critical parameters, with the assigned scores justified by their inherent operational characteristics. The scoring follows a defined scale where a high capability is rated 0.81-1.00, a medium capability 0.61-0.80, a low capability 0.40-0.60, and a score of 0 indicates the parameter is entirely absent The scoring methodology of Table~\ref{tab:relay_capabilities_justified} will be discussed later in this subsection.

Using the weights of different scenarios from Table~\ref{tab:scenario_weights_justified} and the parameter values from Table~\ref{tab:relay_capabilities_justified}, Table~\ref{tab:mcres_computed} presents the computed MCRES values for ARs and TRs. The results show that ARs dominate in scenarios requiring mobility and rapid deployment, such as dynamic battlefields, search and rescue, disaster response, and urban warfare. Conversely, TRs provide higher resilience in missions demanding persistence and concealment, including covert operations, long-endurance surveillance, and fixed-base communications. EW represents a balanced case, where both AR and TR achieve similar performance scores with only a slight difference. Here, the choice depends on the mission profile: ARs are preferable when the scenario requires rapid repositioning, on-the-move spectrum sensing or localization, temporary EW support near maneuvering forces, or the ability to create agile, short-lived protected corridors; TRs are preferable when the scenario requires persistent spectrum monitoring, hardened high-power countermeasures, continuous gateway/anchor services for headquarters or bases, or long-duration anti-jamming operations. Overall, the findings emphasize that the optimal relay choice depends on mission-specific requirements rather than a single universal solution.

Once AR or TR is selected, choose active relays for long-range, high-throughput, and robust links, and RIS relays for low-power, covert, or temporary coverage. Hybrid deployments can combine both for efficiency and resilience.

\begin{algorithm}[t]
\caption{MCRES-Based Relay Selection for Military Scenarios}
\label{alg:mcres_relay_selection}
\begin{algorithmic}[1]
\STATE \textbf{Input:} Scenario weights $w = [w_M, w_J, w_D, w_S, w_C, w_A]$ (Table~\ref{tab:scenario_weights_justified}), 
Relay features $v_r = [M, J, D, S, C, A]$ for each relay $r \in \{\text{AR, TR}\}$ (Table~\ref{tab:relay_capabilities_justified}), and $\delta=0.03$
\STATE \textbf{Output:} Selected relay $r^*$ and recommended deployment type (Active or RIS)

\FOR{each relay $r$}
    \STATE Compute $\text{MCRES}_r$ using (4)
\ENDFOR

\STATE Identify relay $r^* = \arg\max_r \text{MCRES}_r$
\STATE Calculate $\Delta = \mid \text{MCRES}_{\text{AR}} - \text{MCRES}_{\text{TR}} \mid$
\IF{$\Delta \leq \delta$}
    \IF{mission requires mobility, rapid deployment, or temporary EW support}
        \STATE Choose $r^* = \text{AR}$
    \ELSE
        \STATE Choose $r^* = \text{TR}$
    \ENDIF
\ENDIF

\IF{long-range, high-throughput, or robust link needed}
    \STATE Set deployment type $\gets$ Active
\ELSIF{low-power, covert, or temporary coverage required}
    \STATE Set deployment type $\gets$ RIS
\ENDIF
\STATE \textbf{Return:} $r^*$ and deployment type
\end{algorithmic}
\end{algorithm}

\begin{table*}[t]
\caption{MCRES Sensitivity Analysis Across Scenarios and Weight Variations}
\label{tab:mcres_Sensitivity}
\centering
\footnotesize
\begin{tabular}{p{3.2cm} p{6.0cm} p{2.0cm} p{2.0cm} p{2.0cm}}
\hline
\textbf{Scenario} & \textbf{Weight Configuration} & $\mathbf{MCRES_{AR}}$ & $\mathbf{MCRES_{TR}}$ & \textbf{Preferred} \\
\hline
\multicolumn{5}{c}{\textbf{Dynamic Battlefield (Nominal: $w_M=0.30, w_J=0.25, w_D=0.25, w_S=0.20, w_C=0, w_A=0$)}} \\ 
\hline
Nominal & $[0.30, 0.25, 0.25, 0.20, 0, 0]$ & 0.888 & 0.400 & AR \\
$w_M -33\%$ & $[0.20, 0.30, 0.25, 0.20, 0, 0.05]$ & 0.826 & 0.495 & AR \\
$w_M +20\%$ & $[0.36, 0.22, 0.23, 0.19, 0, 0]$ & 0.920 & 0.336 & AR \\
$w_J +33\%$ & $[0.28, 0.33, 0.22, 0.17, 0, 0]$ & 0.863 & 0.490 & AR \\
\hline
\multicolumn{5}{c}{\textbf{Electronic Warfare (Nominal: $w_M=0.10, w_J=0.40, w_D=0, w_S=0.40, w_C=0.10, w_A=0$)}} \\
\hline
Nominal & $[0.10, 0.40, 0, 0.40, 0.10, 0]$ & 0.800 & 0.780 & $\Delta < \delta$ \\
$w_J +15\%$, $w_S -5\%$ & $[0.08, 0.46, 0, 0.38, 0.08, 0]$ & 0.790 & 0.809 & $\Delta < \delta$ \\
Mobility Emphasis & $[0.25, 0.30, 0, 0.30, 0.15, 0]$ & 0.850 & 0.630 & AR \\
\hline
\multicolumn{5}{c}{\textbf{Covert Operations (Nominal: $w_M=0, w_J=0.25, w_D=0, w_S=0.50, w_C=0, w_A=0.25$)}} \\
\hline
Nominal & $[0, 0.25, 0, 0.50, 0, 0.25]$ & 0.713 & 0.850 & TR \\
$w_S -20\%$ & $[0, 0.25, 0, 0.40, 0, 0.35]$ & 0.740 & 0.840 & TR \\
$w_A +20\%$ & $[0, 0.23, 0, 0.45, 0, 0.32]$ & 0.724 & 0.843 & TR \\
Stealth Emphasis & $[0, 0.20, 0, 0.60, 0, 0.20]$ & 0.710 & 0.835 & TR \\
\hline
\multicolumn{5}{c}{\textbf{Search and Rescue (Nominal: $w_M=0.40, w_J=0, w_D=0.40, w_S=0, w_C=0.20, w_A=0$)}} \\
\hline
Nominal & $[0.40, 0, 0.40, 0, 0.20, 0]$ & 1.000 & 0.160 & AR \\
$w_M -25\%$ & $[0.30, 0, 0.45, 0, 0.25, 0]$ & 0.975 & 0.215 & AR \\
Coverage Emphasis & $[0.35, 0, 0.35, 0, 0.30, 0]$ & 0.985 & 0.180 & AR \\
\hline
\end{tabular}
\end{table*}

Environmental and logistical factors are also critical in evaluating UAV relay deployment alongside the MCRES framework. These factors include the operational weather envelope, refuel or recharge requirements, and launch/landing constraints. AR relays typically offer greater mobility but are more sensitive to adverse weather and require periodic energy replenishment, while TR relays provide stable operation in all weather conditions with continuous power supply but are restricted to fixed launch and landing sites. Considering these factors allows planners to determine the feasibility of relay deployment and identify scenarios where mixed AR+TR configurations may be advantageous, such as combining mobile coverage extension with a stable backbone link.\\

The pseudocode of the proposed MCRES-based relay selection framework is shown in Algorithm~\ref{alg:mcres_relay_selection}.

Algorithm~\ref{alg:mcres_relay_selection} provides a structured procedure for selecting the most suitable relay for military scenarios using the MCRES. The algorithm takes as input the mission-specific weights for each parameter and the relay characteristics, computes a weighted sum (MCRES) for each relay, and selects the relay with the highest score. Special rules are included for balanced scenarios, to adjust the choice based on operational priorities. Finally, the algorithm recommends the deployment type—Active or RIS—based on mission requirements such as link robustness, power constraints, or covert operations. This structured approach implements a weighted scoring mechanism consistent with MCDM practices \cite{Triantaphyllou2000}, providing an efficient tool to guide military relay selection.

It is important to note that the small threshold $\delta$ ensures that when $\Delta\mathrm{MCRES}$ falls below this value, the two relay types are treated as operationally equivalent. This accounts for the fact that MCRES scores are based on estimated parameters, so small differences may reflect approximation rather than real capability gaps. From Table~9, most scenarios show clear margins ($|\Delta\mathrm{MCRES}| > 0.10$), while only EW has a narrow difference ($\Delta\mathrm{MCRES} = 0.020$). Therefore, selecting $\delta = 0.03$ is appropriate: it triggers contextual review for the single ambiguous case and allows automatic relay selection for all scenarios with clear preference. A threshold of $\delta = 0.02$ would also be suitable, as it still flags EW for review. However, setting $\delta = 0.01$ would result in an automatic selection of the AR relay type in all scenarios, eliminating the ability to detect borderline cases.

Building on the MCRES framework, the following points detail its implementation: (1) the complexity and input-data assumptions of Algorithm 1, (2) the sensitivity of the MCRES framework to mission priorities, and (3) the scoring methodology used to quantify relay capabilities.

\subsubsection{The complexity and input-data assumptions of Algorithm 1}
Algorithm~1 has $O(1)$ computational complexity for a fixed set of relay types, requiring only simple weighted summations and a single comparison. Its input assumptions are: (1) a normalized weight vector reflecting mission priorities (Table~7); (2) pre‑defined, static capability scores for each relay type (Table~8); (3) a fixed equivalence threshold $\delta$; and (4) mission maps. The algorithm does not rely on real‑time sensor data or iterative optimization, making it suitable for rapid, resource‑constrained decision‑support, but it presumes that weights and capability scores have been accurately calibrated beforehand.

\subsubsection{The sensitivity of the MCRES framework}
To assess the robustness of the proposed MCRES metric, a sensitivity analysis was conducted by varying the mission priority weights across different operational scenarios. The results detailed in Table~\ref{tab:mcres_Sensitivity} validates the robustness of the proposed MCRES metric across a range of mission‑critical operational environments. For the majority of scenarios, the recommended relay type remains stable even under significant variations in priority weights. In the dynamic battlefield scenario, for example, reducing the mobility weight  or increasing jamming resilience by a third does not alter the clear preference for ARs, whose scores consistently exceed those of TRs by a wide margin. Similarly, in search and rescue, ARs retain an overwhelming advantage across all tested weight configurations, underscoring the metric’s reliability in missions where rapid deployment and coverage are paramount.

The sensitivity results highlight a critical feature of the MCRES framework: its explicit handling of operationally equivalent outcomes. This is most evident in the EW scenario. The table shows that under both its nominal weight configuration and a variant prioritizing jamming resilience, the MCRES scores for ARs and TRs are nearly identical, with differences $\Delta = 0.020$ and $\Delta = 0.019$. According to Algorithm 1, when $\Delta \leq \delta$ (the decision threshold), the framework does not automatically select the higher-scoring relay.

The inclusion of a mobility emphasis test case within the EW scenario further illustrates this logic: when mobility is explicitly prioritized, the score difference grows (0.850 vs. 0.630), and the algorithm confidently recommends AR.

\subsubsection{Scoring Methodology}
 The numerical scores presented in Table~\ref{tab:relay_capabilities_justified}  are derived from a structured, evidence-based comparative framework that translates documented operational characteristics into normalized quantitative ratings for systematic analysis. Each of the six mission-critical parameters was evaluated on a standardized $0$--$1.0$ scale, where scores reflect relative capability levels rather than absolute measurements from any single source. This scoring methodology was developed through two complementary processes: first, a comprehensive review of performance data and limitations reported in UAV and terrestrial relay literature; second, alignment with military doctrinal priorities such as agility (JCIDS), spectrum resilience (DoDI~4650.01), and communication reliability (AJP-6).

 For example, the aerial relay's mobility score of $1.00$ is assigned because UAVs possess unrestricted three-dimensional mobility---a consistently highlighted operational advantage in studies on dynamic coverage optimization, adaptive positioning, and threat evasion. Research demonstrates that UAVs can be repositioned in real-time to maintain line-of-sight links, avoid obstructions, and extend network reach in contested environments, fundamentally distinguishing them from static or ground-constrained systems. In contrast, the terrestrial relay receives a mobility score of $0.00$, reflecting its inherent fixed deployment nature and inability to dynamically relocate---a well-documented limitation in surveys of ground-based relay infrastructure.

Similarly, jamming resilience scores illustrate this comparative approach: terrestrial relays are rated $1.00$ due to their ability to leverage grid power, hardened installations, and high-power countermeasures, whereas aerial relays receive $0.75$ because, despite some evasion capabilities, they remain constrained by limited onboard power, aerodynamic design compromises, and heightened susceptibility to directed electronic attack. Deployment speed further exemplifies this methodology: aerial relays score $1.00$ based on documented rapid-launch capabilities and minimal setup requirements, while terrestrial relays score $0.00$ due to their time-intensive installation, site preparation, and logistical footprint.

This synthesized ordinal ranking system enables clear, operationally meaningful comparisons between relay archetypes. By converting qualitative capabilities and documented limitations into a consistent numerical framework, this scoring system supports objective trade-off analysis and facilitates the mission-aware relay selection process formalized in the MCRES framework.

\subsection{Lessons Learned}
The analysis based on the MCRES framework highlights several critical insights for mission planners. First, ARs demonstrate clear advantages in mobility and rapid deployment, making them highly effective in contested and dynamic environments such as battlefield operations and disaster response, where communication nodes must be established quickly under uncertainty. However, these gains come at the cost of limited endurance and greater logistical burden. Second, TRs offer superior sustainability and resilience, benefiting from fixed infrastructure and continuous power, which makes them ideal for long-endurance surveillance, base communications, and persistent monitoring scenarios. Nevertheless, this comes at the expense of flexibility and rapid redeployment. Finally, the results underscore that no single relay type dominates across all operational parameters; instead, hybrid approaches, leveraging the agility of ARs alongside the persistence and energy efficiency of TRs, emerge as a promising pathway toward resilient and adaptable military communication networks.

\section{Challenges and Future Directions} 
\subsection{Challenges}	
UAV-based military relays present several technical challenges that need to be addressed for effective military applications:
\vspace{-1em}
\subsubsection{Limited Endurance and Payload Capacity} 
One of the most pressing challenges for UAVs used as aerial relays in military operations lies in balancing endurance and payload capacity. Limited battery life or fuel restricts flight time and range, making it difficult to sustain persistent ISR or coverage in extended missions. Payload constraints further limit the ability to carry high-power radios, EW modules, or advanced sensors, reducing overall relay effectiveness. While larger UAVs can extend endurance and host heavier payloads, they become more costly, less agile, and more vulnerable to detection—traits that undermine rapid deployment and survivability in contested environments. Thus, the key challenge is achieving a balance between endurance, payload, agility, and cost, ensuring UAV relays remain effective in fast-moving and resource-constrained battlefield scenarios.

\vspace{-1em}
\subsubsection{Complexity in Coordination and Control} 
As the deployment of multiple UAVs becomes more prevalent in military operations, coordinating their movements and maintaining secure, resilient communications presents significant challenges. This complexity is heightened in contested, rapidly changing battlefields, where UAVs must operate under adversarial interference, EW, and time-critical mission demands. Effective swarm operation requires advanced algorithms for collision avoidance, dynamic flight-path optimization, distributed decision-making, secure, and low-latency C2 links. Additionally, UAVs must adapt in real time to shifting enemy positions, evolving mission objectives, and emerging threats such as surface-to-air systems or cyber attacks. Survivability also demands threat detection, evasive maneuvers, and signature management. Robust coordination frameworks are therefore essential to ensure autonomy, operational safety, resilient networking, and mission success in highly contested military environments.

\subsubsection{Susceptibility to Electronic Warfare} 
As discussed throughout the paper, military UAVs are highly vulnerable to EW threats, including jamming, spoofing, and cyberattacks, which can disrupt communications, misdirect platforms, or compromise missions. Maintaining secure, resilient links and protecting sensitive data is critical in contested environments. Equipping UAVs with anti-jamming technologies, encrypted communications, frequency agility, and robust protocols is essential, and a strong, adaptive algorithm to counter EW attacks is crucial. These measures enhance survivability but also increase payload and system complexity, which must be balanced with endurance, agility, and overall mission efficiency.

\subsubsection{Environmental Factors} 
Military UAVs are highly sensitive to environmental factors that can degrade performance. Adverse weather conditions—such as high winds, rain, snow, sandstorms, or extreme temperatures—can compromise stability, sensor functionality, and propulsion efficiency. In military operations, UAVs often operate in unpredictable and harsh environments, including combat zones, disaster areas, deserts, or mountainous terrain. Designing UAVs to withstand these conditions requires robust materials, specialized sensors, and propulsion systems capable of maintaining stability and maneuverability under environmental stress. Additionally, UAVs must sustain communication and navigation in areas where line-of-sight is limited, such as dense forests, urban areas, or rugged terrain, ensuring reliable mission execution in challenging operational contexts.

\subsubsection{Avoiding Radar Detection} 
A critical challenge in military UAV operations is avoiding detection by enemy radar systems. UAVs conducting surveillance, reconnaissance, or strike-support missions are particularly at risk of being located and targeted. The RCS of a UAV directly influences its detectability, making RCS reduction essential for mission survivability. Effective stealth strategies include the use of radar-absorbent materials, airframe designs that minimize radar reflections, and tactical flight maneuvers such as low-altitude operations or terrain masking. The key challenge is integrating these stealth techniques while preserving operational performance, including communication links, endurance, maneuverability, and payload capacity, to ensure UAVs can complete missions effectively in contested airspace.

\subsubsection{Lightweight RIS Development}	 Integrating RIS onto military UAVs poses a significant challenge, particularly for smaller platforms where additional weight can degrade flight performance, stability, and endurance. Lightweight, high-performance RIS technologies are critical for UAVs operating in tactical environments that demand rapid deployment, high mobility, and prolonged mission duration. The challenge lies in designing RIS components that are compact, energy-efficient, and low-mass while still enhancing communication capabilities, such as extending secure line-of-sight links, supporting resilient C2 networks, and improving signal propagation in complex or contested terrains.

\subsubsection{Trajectory design and power Optimization}	  
Optimizing the trajectory and power usage of UAV-based military relays is a highly complex task, constrained by multiple operational factors. Key challenges include balancing flight stability, energy efficiency, endurance, and reliable signal coverage in contested environments. Effective trajectory and placement algorithms must consider UAV flight dynamics, payload limitations, environmental conditions, and mission-specific requirements. Different operations impose varying priorities: wide-area surveillance missions demand maximum coverage and strong communication links, whereas covert or contested operations require minimal signal exposure, stealth, and survivability. Additionally, UAV relays must adapt in real time to dynamic battlefield conditions, such as shifting enemy positions, obstacles, and sudden mission changes. Developing algorithms capable of dynamically optimizing UAV trajectories, and power allocation under these constraints is critical for ensuring resilient, efficient, and mission-effective aerial relay networks in military operations.
   
\subsection{Future Directions} 
\subsubsection{Enhanced Power Management and Lightweight Materials} Research is needed to develop advanced power management systems and incorporate lightweight materials that extend battery life and increase payload capacity without significantly increasing the UAV's weight. Optimal flight path planning that conserves energy is crucial for improving endurance and operational efficiency. Additionally, this should be paired with novel lightweight UAV designs that do not compromise flight stability or payload constraints. Furthermore, advanced techniques like beamforming, optimized multi-antenna systems, frequency control, energy harvesting, and adaptive power management can be used to improve charging efficiency \cite{8579209}.
\subsubsection{Advanced Design Algorithms} Developing robust coordination algorithms for UAV swarms remains a significant challenge, particularly for large-scale military operations. These algorithms must efficiently manage dynamic environments and ensure optimal 3D placement of UAVs, minimizing collision risks while maximizing network performance. Moreover, there is a need for joint optimization of UAV deployment and bandwidth allocation to ensure low-latency communications, as well as obstacle-aware coordination for maximizing wireless coverage.
\subsubsection{Security and Privacy} The integration of UAV-enabled networks in military operations requires addressing critical security challenges to ensure reliable and resilient functionality. Research should focus on mitigating physical threats that target hardware components and countering cyber vulnerabilities such as jamming, spoofing, and data fabrication, which can compromise system performance. Securing communication, control, feedback, sensing, and storage mechanisms is crucial for robust operation. Furthermore, the widespread deployment of UAV networks introduces significant privacy concerns, particularly regarding the misuse of sensitive data, including geographic locations and personal information. Developing advanced solutions to safeguard privacy and facilitate secure data sharing is essential to support the growing reliance on UAV networks.
\subsubsection{Resilient Designs for Various Environmental Conditions} UAVs often face diverse environmental challenges such as high winds, precipitation, and temperature extremes, which can significantly impact their performance. Developing an A2G channel model that considers environmental factors like temperature, wind, foliage, and urban landscapes is important. Then, based on this model, research is required to develop resilient design solutions that ensure operational effectiveness in adverse conditions. Additionally,  this could involve creating materials and structural designs that withstand extreme weather or environmental conditions while maintaining communication reliability.
\subsubsection{UAV Deployment and 3D Placement Optimization} 
In military applications, optimizing the 3D placement of UAVs is crucial not only for enhancing network performance but also for ensuring operational safety. Strategic deployment must account for potential threats, positioning UAVs in secure locations away from enemy detection and attacks. This is especially vital when UAVs operate alongside terrestrial networks, requiring interference management and precise positioning to minimize detection risks. For high-frequency bands like mmWave, where signals are highly susceptible to blockages, careful placement is necessary to maintain robust coverage \cite{khan2021role}. Additionally, optimizing UAV altitude and trajectory must consider A2G channel characteristics, flight endurance, and energy efficiency, ensuring both secure and reliable communication.

\begin{table*}[htbp]
\caption{Research Directions, Challenges, and Analytical Methods for UAV-based Aerial Relay Systems in Military Applications}
\centering
\begin{tabular}{@{}p{3.2cm} p{7cm} p{4.5cm} p{1.8 cm}@{}}
\toprule
\textbf{Research Direction} & \textbf{Challenges} & \textbf{Approaches} & \textbf{Refs.} \\ 
\midrule
Enhanced Power Management & Balancing endurance, payload, and lightweight RIS design & Energy-efficient planning, harvesting, lightweight materials & \cite{9902968,10032196,10208231,8943430,8561238,8579209} \\ 
\midrule
Advanced Design Algorithms & Coordinating large fleets, collision avoidance, low-latency C2 & RL, game theory, multi-agent, distributed optimization & \cite{9834117} \\ 
\midrule
Security and Privacy & Countering jamming, spoofing, cyber-attacks & Crypto, anti-jamming, frequency hopping, coding & \cite{10462233,9454372,9687318,8255739,Altawy2016} \\ 
\midrule
Resilient Designs & Survivability under weather, terrain, and NLoS & Channel modeling, robust structures, materials & \cite{10069507} \\ 
\midrule
UAV Deployment & Secure placement, coverage vs. stealth, blockage mitigation & RL, game theory, placement optimization & \cite{10478734,9739696,10122731,khan2021role} \\ 
\midrule
Trajectory Optimization & Stealthy, energy-efficient, threat-aware paths & RL, decentralized control, motion planning & \cite{9767553,9826431,8676325,9826431,10586977,10533208} \\ 
\midrule 
Network Integration & Terrestrial-satellite integration, spectrum coordination & Cross-domain models, scalability analysis & \cite{10612249,10499205} \\ 
\midrule
Resource Management & Bandwidth, power, and spectrum under EW & Resource allocation, dynamic programming & \cite{10533208,8708975,9169676,10584067} \\ 
\midrule
Autonomy and Swarms & Adaptive swarms, real-time threat response & AI/ML, swarm intelligence, adaptive control & \cite{9834117,10430396} \\ 
\midrule
Performance Analysis & Military metrics under mobility and interference & Modeling, queuing, scheduling analysis & \cite{7412759,7967745} \\ 
\midrule
Digital Twins & Real-time replicas for rehearsal and prediction & Simulation, AI prediction, edge analytics & \cite{10190734,10045049,10371218} \\ 
\midrule
Other Topics & RIS on small UAVs, stealth, LPI/LPD tactics & EM modeling, stealth design, RCS reduction & --- \\
\bottomrule
\end{tabular}
\label{tab:aerial_relays}
\end{table*}

\subsubsection{Secure UAV Trajectory Optimization}
The problem of UAV trajectory design has attracted significant research attention in recent years. Due to its non-convex nature, obtaining a global optimum is challenging, and conventional optimization methods generally require complete knowledge of the network environment, which is rarely available or difficult to estimate \cite{9826431}. To address this, reinforcement learning (RL) has emerged as an effective approach for joint trajectory and power optimization in multi-UAV networks, as it can learn and adapt from experience without relying on full environmental knowledge \cite{9767553,9826431,8676325}. However, most of these studies focus on civilian applications where threats are modeled as static obstacles in the environment. 

In military contexts, UAV trajectory design must address highly contested war zones with mobile and adversarial threats. Research should focus on developing threat-aware and stealth-preserving flight paths. Trajectories must jointly consider mobility, energy efficiency, and secure communication under partial and uncertain information. Unlike most civilian studies that rely on offline-trained algorithms, online learning and adaptation are crucial in military applications to respond to rapidly changing threats and dynamic environments. Future directions include resilience-aware reinforcement learning, prediction of mobile threats, and cooperative swarm maneuvers. Decentralized control and real-time adaptation are also critical. Finally, communication-aware path planning with low probability of detection is needed to reduce risks in battlefield conditions.
\subsubsection{Seamless Network Integration and Scalability} 
For military operations, the seamless integration of UAV-based relays with terrestrial command networks and satellite communication assets is critical to ensure unified, secure, and resilient connectivity across diverse theaters of operation. Future research should emphasize mission-driven network planning and integration strategies that enable rapid deployment and scalable support for joint and coalition forces. Moreover, detailed analysis of signaling overhead, spectrum coordination, and resource allocation mechanisms is essential to guarantee scalability, operational efficiency, and reliable performance under contested and high-mobility conditions.
\subsubsection{Resource Management in UAV Networks} 
Effective resource management remains a critical challenge in UAV-enabled military communications, where bandwidth, energy, flight endurance, and transmit power must be allocated under strict mission and operational constraints. Future solutions should account for the dynamic and unpredictable nature of battlefield environments, including UAV mobility, hostile interference, varying traffic loads, and line-of-sight limitations in complex terrain. Additionally, mission-aware frequency planning and spectrum sharing between aerial relays, terrestrial infrastructure, and satellite links are essential to maximize efficiency, ensure resilience against jamming, and sustain reliable command, control, and intelligence dissemination in contested domains.
\subsubsection{Increased Autonomy and Swarm Tactics}  
Future research should advance the development of autonomous UAV swarms capable of conducting complex military missions with minimal operator oversight. Emphasis should be placed on scalable and adaptive swarm management techniques that enable coordinated maneuvering, distributed sensing, and cooperative strike or support operations. The integration of AI and ML can empower swarms to adapt dynamically to battlefield conditions, anticipate threats, and make mission-critical decisions in real time. AI-driven algorithms can optimize formation control, task allocation, and survivability under EW and contested spectrum environments, while ML models can enhance communication reliability by predicting disruptions and autonomously adjusting routing and coordination strategies. Such advancements will strengthen swarm intelligence and ensure resilient, mission-ready performance across diverse operational theaters.
\subsubsection{Performance Analysis}  
Future research should conduct comprehensive performance evaluations of UAV-based aerial relays to optimize next-generation military networks. Priority areas include developing analytical models to assess coverage, spectral efficiency, and energy efficiency in integrated aerial–terrestrial–satellite architectures. In addition, the impact of UAV mobility on mission-critical metrics such as throughput, latency, and endurance must be quantified to ensure reliable C2 and timely ISR data dissemination. Evaluating dynamic scheduling and adaptive relay selection under contested and high-mobility conditions will also be essential for enhancing robustness, resilience, and overall system performance in operational environments.
\subsubsection{Digital Twins for Military UAV Relays}
Digital twins  for military UAVs provide real-time virtual replicas of UAVs and their operational environments, supporting mission rehearsal, predictive and what-if analyses, and strategic decision-making. They can enable life cycle optimization of aerial relays, improving deployment planning, maintenance, and operational readiness. Future research should address scalability and real-time performance for large UAV swarms, data synchronization and fidelity across heterogeneous sensors, and robust security and interoperability frameworks to protect sensitive mission data. Additional work is needed for multi-domain integration with land and maritime systems and for developing reliable AI-driven autonomous capabilities under high-stakes battlefield conditions. These advances will enhance UAV relay network resilience, efficiency, and operational effectiveness.
\subsubsection{Other Interesting Topics}  
Beyond the aforementioned prospects, several additional research directions are particularly relevant for military applications. One important topic is the development of lightweight, high-performance RIS units designed for smaller UAV platforms, enabling improved signal coverage while maintaining flight dynamics and endurance. Another pressing challenge concerns minimizing the radar detectability of UAV-based relays, which is critical for survivability in contested airspaces. Addressing this issue requires advancements in stealth technologies, including the use of radar-absorbent materials, airframe designs optimized to reduce radar cross-section, and adaptive flight strategies that support LPI/LPD. Such innovations would enable UAV relays to operate covertly in hostile environments, thereby enhancing their operational effectiveness in combat and intelligence missions.

 Table~\ref{tab:aerial_relays} summarizes the research directions, challenges, and analytical methods for UAV-based aerial relay systems.

\subsection{Lessons Learned}  
The analysis of challenges and future directions for UAV-based aerial relays reveals several foundational imperatives. The persistent trade-off between platform agility and operational endurance underscores the necessity for breakthroughs in power management and lightweight materials. Furthermore, the acute vulnerability to EW dictates that resilience must be a core design axiom, integrated through hardened protocols and decentralized, swarm-based architectures to eliminate single points of failure. Consequently, operational effectiveness in contested spectrums will be governed by advanced autonomy, leveraging machine intelligence for real-time threat response beyond human reaction times. Ultimately, transcending these hurdles demands a cross-disciplinary paradigm, fusing communications theory with materials science and computer engineering to co-optimize stealth, durability, and intelligent coordination for future military networks.

\section{Conclusion}	
This survey demonstrates that UAV-mounted aerial relays are transformative enablers for future military communications, offering decisive advantages in mobility, adaptability, and coverage over traditional terrestrial relays. Their ability to deliver resilient links in contested and infrastructure-denied environments makes them indispensable for modern C2 and CIS operations.

To bridge the gap between potential and deployment, we proposed the MCRES and its decision algorithm. By integrating six weighted attributes into a linear metric, MCRES provides a transparent, repeatable, and mission-specific decision tool. The framework is fully consistent with MCDM practices and aligned with JCIDS, CIS, and AJP-6, ensuring doctrinal interoperability and operational relevance.

Future advances in energy endurance, EW resilience, and swarm autonomy will be essential to overcome current limitations. With these developments, ARs are poised to become a cornerstone of next-generation military networks, delivering assured, secure, and adaptive connectivity in the most demanding operational theaters.

\footnotesize    
\bibliography{UAV_Survey_Ref}
\bibliographystyle{ieeetr}

\end{document}